\documentclass{SciPost}
\usepackage{amsmath, amssymb}
\usepackage{graphicx, nicefrac, textcomp}
\usepackage{color}
\usepackage{dcolumn}
\usepackage[utf8]{inputenc}
\usepackage{float}
\usepackage{dsfont}
\usepackage{bbold}
\usepackage{svg}
\usepackage[normalem]{ulem}
\usepackage[version=4]{mhchem}
\usepackage[toc,page]{appendix}
\usepackage{changes}
\usepackage{comment}
\usepackage{placeins}
\usepackage{tcolorbox}
\usepackage{etoolbox}
\usepackage{fancyvrb}
\usepackage[shortlabels]{enumitem}

\usepackage[frozencache,cachedir=.]{minted}

\hypersetup{
    colorlinks,
    linkcolor={red!50!black},
    citecolor={blue!50!black},
    urlcolor={blue!80!black}
}

\usepackage[bitstream-charter]{mathdesign}
\DeclareSymbolFont{usualmathcal}{OMS}{cmsy}{m}{n}
\DeclareSymbolFontAlphabet{\mathcal}{usualmathcal}

\fancypagestyle{SPstyle}{
\fancyhf{}
\lhead{\colorbox{scipostblue}{\bf \color{white} ~SciPost Physics Codebases }}
\rhead{{\bf \color{scipostdeepblue} ~Submission }}

\fancyfoot[C]{\textbf{\thepage}}
}

\BeforeBeginEnvironment{minted}{
\begin{tcolorbox}[colframe=white!75!blue]
}
\AfterEndEnvironment{minted}{\end{tcolorbox}}%

\mathchardef\up="0222
\mathchardef\dn="0223

\newcommand{\ph}[1]{{\color{magenta}[Ph.: #1]}}

\begin{document}
\pagestyle{SPstyle}

\begin{center}{\Large \color{scipostdeepblue}{\textbf{SOLAX: A Python solver for fermionic quantum systems with neural network support}}}\end{center}

\begin{center}
Louis Thirion\textsuperscript{1},
Philipp Hansmann\textsuperscript{1},
Pavlo Bilous\textsuperscript{2*}
\end{center}

\begin{center}
{\bf 1} Department of Physics, Friedrich-Alexander-Universit\"at Erlangen-N\"urnberg, 91058 Erlangen, Germany
\\
{\bf 2} Max Planck Institute for the Science of Light, Staudtstraße 2, 91058 Erlangen, Germany
\\
*pavlo.bilous@mpl.mpg.de
\end{center}

\begin{center}
\today
\end{center}

\section*{Abstract}

Numerical modeling of fermionic many-body quantum systems presents similar challenges across various research domains, necessitating universal tools, including state-of-the-art machine learning techniques. Here, we introduce SOLAX, a Python library designed to compute and analyze fermionic quantum systems using the formalism of second quantization. SOLAX provides a modular framework for constructing and manipulating basis sets, quantum states, and operators, facilitating the simulation of electronic structures and determining many-body quantum states in finite-size Hilbert spaces. The library integrates machine learning capabilities to mitigate the exponential growth of Hilbert space dimensions in large quantum clusters. The core low-level functionalities are implemented using the recently developed Python library JAX. Demonstrated through its application to the Single Impurity Anderson Model, SOLAX offers a flexible and powerful tool for researchers addressing the challenges of many-body quantum systems across a broad spectrum of fields, including atomic physics, quantum chemistry, and condensed matter physics.
\vspace{\baselineskip}

\noindent\textcolor{white!90!black}{%
\fbox{\parbox{0.975\linewidth}{%
\textcolor{white!40!black}{\begin{tabular}{lr}%
  \begin{minipage}{0.6\textwidth}%
    {\small Copyright attribution to authors. \newline
    This work is a submission to SciPost Physics Codebases. \newline
    License information to appear upon publication. \newline
    Publication information to appear upon publication.}
  \end{minipage} & \begin{minipage}{0.4\textwidth}
    {\small Received Date \newline Accepted Date \newline Published Date}%
  \end{minipage}
\end{tabular}}
}}
}

\tableofcontents

\section{Introduction}

Accurate numerical modeling of fermionic quantum many-body systems presents an essential challenge across many research domains. In atomic physics, for example, precise knowledge of electronic energy levels is indispensable for the development of atomic frequency standards, the understanding of astrophysical spectra, and the search for phenomena beyond the Standard Model~\cite{Fischer_Review_JPhysBAtMolOptPhys_2016}. In particular, the promising yet scarcely explored domain of highly charged ions lacks experimental data and requires extensive computational support~\cite{RevModPhys_HCI_2018}. In quantum chemistry, the pursuit of highly accurate electronic structure calculations, such as those achieved through full configuration interaction (full CI) methods~\cite{ROSSI1999,Huron1973,Greer1998, Garniron2018,Tubman2020}, is crucial for the accurate determination of molecular properties and the prediction of chemical reactivity. In condensed matter physics, quantitative research on low-energy effective Hamiltonians such as the Hubbard model~\cite{HubbardModel} and its derivatives, supports the qualitative understanding of microscopic mechanisms underlying phenomena like unconventional superconductivity in cuprates~\cite{Cuprates1,Cuprates2}, iron pnictides~\cite{pnictides1}, and nickelates~\cite{nickelates1,nickelates2}. While the core motivations and goals in all these diverse research areas are often completely different, the computational challenges are similar and the most challenging tasks are often identical from the technical point of view. Alongside methodological developments, advanced simulation codes for quantum many-body systems are therefore essential for scientific progress across a broad spectrum of fields.

Despite these advances, many challenges still demand computational efforts that exceed the capabilities of even the most efficient codes and/or the available computational resources. In such cases, machine learning techniques can be applied to reduce the complexity of the calculation without compromising the accuracy. Possible approaches based on a neural network (NN) were demonstrated in Refs.~\cite{MLGRASP, Bilous_dlw_gen_2024} for problems in computational atomic physics, and in Ref.~\cite{Bilous2024_SIAM} in a more general context of fermionic systems requiring large expansions of the wave function in the basis of Slater determinants. In the latter work, the machine learning functionality based on the TensorFlow library~\cite{TensorFlow2015} was interfaced with the Quanty CI code~\cite{Lu2014}. Following the successful ``proof of principle'' in Ref.~\cite{Bilous2024_SIAM}, we saw the need to implement an integrated Python library rather than interfacing existing codes. Here we present the resulting SOLAX package.

SOLAX is a comprehensive NN-boosted Python library designed for the study of fermionic quantum many-body systems. Within the standard quantum many-body formalism of second quantization, SOLAX provides a framework for constructing and solving quantum cluster problems. The SOLAX package offers a versatile set of tools to efficiently encode and manipulate basis sets, quantum states, and operators, enabling users to simulate and explore the structure of quantum many-body systems. The library supports the accurate determination of many-body quantum states in finite-size Hilbert spaces. Beyond its core functionalities, it includes built-in machine learning tools. Specifically, SOLAX addresses the exponential growth of Hilbert space dimensions in large quantum clusters: When full diagonalization becomes computationally infeasible, a NN classifier can be employed to approximate the solution through efficient basis optimization. The SOLAX library has already been successfully applied to the study of molecules such as N$_2$~\cite{Schmerwitz2024_N2}. The NN algorithm presented in this article was first demonstrated in Ref.~\cite{Bilous2024_SIAM} in its application to the Single Impurity Anderson Model (SIAM).

Typically, selection methods for basis states were developed in Fortran or similar languages, see e.g. Ref.~\cite{QuantumPackage2.0_2023}. This makes it difficult to include NNs in the code, since the standard libraries for NNs are primarily provided within the Python ecosystem. We implemented SOLAX directly in Python with the core low-level functionalities based on the JAX library recently developed at Google~\cite{jax2018github}. JAX offers highly efficient GPU-accelerated mechanisms to manipulate data organized as arrays, similar to those from the well-known NumPy library~\cite{NumPy}. This allows for a seamless transition of data to and from the NumPy format, which is the main method for storing data in SOLAX. Importantly, JAX was initially designed for high-performance machine learning research and offers powerful capabilities to leverage NNs. We would like to highlight the recently developed JAX-based NetKet package~\cite{NetKet3_10.21468/SciPostPhysCodeb.7, NetKet3.4_10.21468/SciPostPhysCodeb.7-r3.4}, built around the concept of neural quantum states, i.e., NNs that approximately encode states of quantum many-body systems~\cite{Carleo2017}. Here, we follow a different approach from Ref.~\cite{Bilous2024_SIAM}, using a NN classifier to perform selection of important basis states. To the best of our knowledge, SOLAX is the first integrated JAX-based implementation of such a NN-supported approach.


We assume the reader of this article to be familiar with the Python programming language and the libraries NumPy~\cite{NumPy} and SciPy~\cite{scipy} which are both extensively used in scientific programming. The necessary information on the tools from the JAX ecosystem will be provided as they are used. We do not assume that the reader has experience with machine learning using NNs and provide an introduction to the basic NN concepts relevant for this work. For more information on NNs we refer to the classical work~\cite{Goodfellow2016}. Machine learning from the general (probabilistic) perspective is discussed in depth in the comprehensive source~\cite{murphy_ml1, murphy_ml2}. For a practical introduction to machine learning (including neural networks using TensorFlow), we recommend the book~\cite{HandsOnML}.

The article is structured as follows. In Section~\ref{sec:solver}, we showcase the core functionality of SOLAX for solving the fermionic many-body problem and provide an exemplary computation for SIAM. Section~\ref{sec:nn_support} presents the built-in SOLAX tools for NN-assisted computations and their application to SIAM along with computational time benchmarks. In Section~\ref{sec:save_load}, we describe the mechanisms for saving and loading SOLAX objects, as well as reproducing SOLAX computations. The article closes with the conclusions and outlook in Section~\ref{sec:conclusions}.

\subsection{Code availability and dependencies}
The SOLAX code can be cloned directly from the GitHub repository~\cite{solax_repo}, where we also provide Jupyter notebooks with the code snippets shown in this article. Apart from packages from the standard Python library available without separate installation, SOLAX employs the following third-party libraries: NumPy~\cite{NumPy}, Pandas~\cite{pandas}, SciPy~\cite{scipy}, JAX~\cite{jax2018github}, FLAX~\cite{flax2020github}, and Orbax~\cite{orbax} (the latter 3 libraries belong to the JAX ecosystem). The versions of Python and the listed packages used in SOLAX at the moment of the present publication are summarized in Table~\ref{tab:versions}. The user is required to perform the necessary installations before using SOLAX. For installing JAX, we suggest to follow the instructions on the webpage~\cite{jax_install}, where different installation aspects are addressed. For leveraging GPU acceleration, a GPU-capable version of JAX must be installed. The current version of SOLAX includes a possibility to perform some computations  in parallel on multiple GPUs. Note, however, that this functionality is still under development, and is considered here as an experimental feature. We also note that compatibility of the presented SOLAX version with newer versions of the listed packages cannot be guaranteed, especially for the libraries from the JAX ecosystem which are still under extensive development. In further SOLAX versions we plan to take into account the development progress for the dependencies.
\begin{table}[ht!]
\begin{center}
\begin{tabular}{ccccccc}
Python & NumPy & Pandas & SciPy & JAX & FLAX & Orbax \\\hline
3.10.9 & 1.26.1 & 1.5.3 & 1.10.0 & 0.4.30 & 0.8.5 & 0.1.9
\end{tabular}
\caption{The versions of Python and the third-party libraries used in SOLAX at the moment of the present publication.\label{tab:versions}}
\end{center}
\end{table}

\section{Solver for fermionic quantum systems\label{sec:solver}}

The core functionality of the SOLAX package consists of encoding and solving eigenvalue equations for fermionic quantum many-body systems. Fully antisymmetric Slater determinants serve as the basis for many-body Hilbert spaces of fermionic wave functions. The occupation number representation of a Slater determinant is a binary string of zeroes ``0'' and ones ``1'' (adhering to the Pauli exclusion principle). As the full set of Slater determinants forms a complete basis on a given Hilbert space, any many-body quantum state in this space can be expanded using this basis. SOLAX follows this paradigm in the representation of quantum states. Operators within the SOLAX package are expressed in terms of creation ($\hat{a}^\dagger_i$) and annihilation ($\hat{a}_i$) operators which act on basis Slater determinants or quantum states represented as linear combinations of Slater determinants. These ladder operators are indexed by single-particle quantum numbers $i$ which indicate the positions in the occupation-number string on which they act. In this formalism, a standard Hamilton operator can be decomposed into single-particle, two-particle, and more generally, $n$-particle operators. Each of these terms in the Hamiltonian is expressed as a sum of ladder-operator products containing the corresponding number of creation and annihilation operators. 

In the following, we describe how Slater determinant bases, many-body quantum states, operators, and their matrix representation can be defined and manipulated in SOLAX. We first introduce the main components implemented in SOLAX as Python classes: \verb+Basis+, \verb+State+, \verb+OperatorTerm+, \verb+Operator+, and \verb+OperatorMatrix+. Subsequently, we demonstrate these fundamental tools with the help of an example by finding the ground state of the paradigmatic Single Impurity Anderson Model (SIAM). The data storage and processing for the presented classes are primarily based on NumPy arrays~\cite{NumPy}, which also facilitate convenient interaction with the user. Internally, SOLAX enhances operations on 2D NumPy arrays using the Pandas library~\cite{pandas}. This powerful data analysis tool allows us to leverage hash maps and significantly speed up operations in which search in 2D NumPy arrays is needed. At the same time, the central quantum solver operations like e.g. acting with operators on quantum states are implemented in JAX~\cite{jax2018github} and support automatic GPU acceleration.

\subsection{Basis}
We begin by introducing the \verb+Basis+ class in SOLAX, which facilitates efficient manipulation of basis sets composed of Slater determinants. To utilize SOLAX and NumPy in a Python environment, the libraries are imported as follows:

\begin{minted}{python}
import solax as sx
import numpy as np
\end{minted}

\subsubsection{Object construction}

A \verb+Basis+ instance is created from a collection of strings, each representing a Slater determinant in the occupation number representation (note that when initializing a \verb+Basis+ with a single Slater Determinant, the latter still must be included in a Python collection, e.g. a \verb+list+). All determinants (strings) must have the same length corresponding to the number of the underlying single-particle degrees of freedom which we refer to from here on as ``spin-orbitals''. For example, a \verb+Basis+ object \verb+basis+ with all possible Slater determinants for two electrons occupying four spin-orbitals is constructed as follows.
\begin{minted}{python}
basis = sx.Basis(["1100", "1010", "1001", "0110", "0101", "0011"])
\end{minted}
\noindent SOLAX treats the received strings as binary representations of 8-bit integer numbers (in general a few for each determinant) and stores these efficiently in a 2D NumPy array. The length of each Slater determinant (i.e. the total number of 0s and 1s) in a \verb+Basis+ object can be accessed using the read-only \verb+bitlen+ property:
\begin{minted}{python}
print(basis.bitlen)
\end{minted}
\begin{verbatim}
    4
\end{verbatim}
As will be seen in the following, \verb+Basis+ objects behave similarly to Python sets in many contexts. In particular, if repeated determinants are passed to the class constructor, these are automatically discarded.


\subsubsection{Conversion to a Python string and printing}

The user can convert a \verb+Basis+ to a Python string and print it in order to see the stored determinants in the usual format:
\begin{minted}{python}
print(basis)
\end{minted}
\begin{verbatim}
    1100
    1010
    1001
    0110
    0101
    ...
\end{verbatim}
Note that only the first five (the default value of the print limit) determinants are reflected in the printed string. The value of this limit can be changed using the context manager \verb+dets_printing_limit+:
\begin{minted}{python}
with sx.dets_printing_limit(10):
    print(basis)
\end{minted}
\begin{verbatim}
    1100
    1010
    1001
    0110
    0101
    0011
\end{verbatim}
The user provides here the maximal number of printed determinants (in this case $10$) or \verb+None+ if no limit is to be used.

\subsubsection{Length, indexing, and slicing}

\paragraph{A. Basis as a Python sequence.} The \verb+Basis+ class implements the Python sequence protocol, i.e. its objects have length and can be indexed and sliced in the standard way. The length is the number of the contained Slater determinants, and can be obtained with the Python built-in \verb+len+ function:
\begin{minted}{python}
print(len(basis))
\end{minted}
\begin{verbatim}
    6
\end{verbatim}
Indexing and slicing picks the Slater determinants at the corresponding positions and returns a new \verb+Basis+ object:
\begin{minted}{python}
print(basis[0])
\end{minted}
\begin{verbatim}
    1100
\end{verbatim}
\begin{minted}{python}
print(basis[0:3])
\end{minted}
\begin{verbatim}
    1100
    1010
    1001
\end{verbatim}
Note that in contrast to the typical behavior of sequences, \verb+basis[0]+ is again of type \verb+Basis+. However, this is natural for this particular class and still complies with the Python sequence interface.

\paragraph{B. ``Fancy'' and boolean indexing.} On top of the standard Python sequence functionality, the \verb+Basis+ class supports the NumPy-style ``fancy'' and boolean indexing~\cite{python_for_da_book}. That is, \verb+Basis+ objects can be indexed in two additional ways: using a list on indices and with a boolean mask, respectively
\begin{minted}{python}
print(basis[0, 1, 4])
\end{minted}
\begin{verbatim}
    1100
    1010
    0101
\end{verbatim}
\begin{minted}{python}
print(basis[True, True, False, False, True, False])
\end{minted}
\begin{verbatim}
    1100
    1010
    0101
\end{verbatim}
As for NumPy arrays, the boolean mask here must have the same length as the sequence itself.

\paragraph{Important Note!} There is a crucial difference between the indexing mechanisms shown in (A) and (B), which is inherited from NumPy. In NumPy, standard Python indexing and slicing do not create a new array but instead reference a sub-array of the original array. In contrast, fancy or boolean indexing does create a new array in memory. This behavior extends to the \verb+Basis+ class, as it is based on NumPy (like most classes in SOLAX). While a new \verb+Basis+ object is created regardless, this does not necessarily hold for the underlying NumPy arrays. Therefore, for efficient memory usage, the indexing method described in (A) should be preferred.

\subsubsection{Set operations}

From the operational perspective, \verb+Basis+ objects behave similarly to Python sets. However, they do not fully implement the standard Python set interface. The reason for this is twofold: (1) not all set operations are necessary for SOLAX applications, and (2) the standard Python set nomenclature can be somewhat confusing in this context. Below, we describe the operations with \verb+Basis+ objects as implemented in SOLAX.

\paragraph{A. Equality relation.} As for Python sets, the equality relation \verb+==+ disregards the order of the Slater determinants in the compared \verb+Basis+ instances. For example, we compare the created \verb+basis+ object with its reverse:
\begin{minted}{python}
print(basis == basis[::-1])
\end{minted}
\begin{verbatim}
    True
\end{verbatim}

\paragraph{B. Addition  (set union).} Objects of the \verb+Basis+ class can be added (unified like sets) using the \verb|+| operator. Note that the analogous operator for Python sets is \verb+|+. As an example, consider two \verb+Basis+ objects obtained from \verb+basis+ by selecting only Slater determinants at even and odd positions, respectively:
\begin{minted}{python}
basis_even = basis[::2]
basis_odd = basis[1::2]
\end{minted}
\noindent As expected, their addition equals the initial \verb+basis+:
\begin{minted}{python}
print(basis_even + basis_odd == basis)
\end{minted}
\begin{verbatim}
    True
\end{verbatim}
Repeated determinants are automatically excluded from the resulting \verb+Basis+. For example, unification of \verb+basis+ with itself gives again \verb+basis+ since no new determinants are added:
\begin{minted}{python}
print(basis + basis == basis)
\end{minted}
\begin{verbatim}
    True
\end{verbatim}

\paragraph{C. Set difference.} \verb+Basis+ objects can be subtracted like sets. This operation is bound to the operator \verb+%+ instead of \verb+-+ as for Python sets. For instance, for the objects \verb+basis+, \verb+basis_even+ and \verb+basis_odd+ introduced above, we have:
\begin{minted}{python}
print(basis % basis_even == basis_odd)
\end{minted}
\begin{verbatim}
    True
\end{verbatim}
Note that both for the \verb|+| and the \verb+%+ operator, the Slater determinants in both operands must have the same length in the sense of the total number of 0s and 1s (which can be accessed using the \verb+bitlen+ property).

\subsection{State}

We proceed with the \verb+State+ class in SOLAX. While the \verb+Basis+ class incorporates sets of Slater determinants, the \verb+State+ class also includes real or complex coefficients assigned to each determinant. As such, a \verb+State+ object represents a quantum state expanded in the basis of Slater determinants. SOLAX supports linear algebra operations on \verb+State+ objects including their scalar product, thereby implementing the structure of a Hilbert space.

\subsubsection{Fundamentals}

Here we demonstrate the basic usage of the \verb+State+ class. As will be seen, many aspects are similar to those of the \verb+Basis+ class. A \verb+State+ instance is created from a \verb+Basis+ instance and a NumPy array of associated coefficients:
\begin{minted}{python}
state = sx.State(basis, np.ones(6))
\end{minted}
\noindent Here we used the \verb+basis+ object created in the previous section and a NumPy array with real numbers all equal 1. Note that states as represented by the SOLAX \verb+State+ class can be unnormalized, e.g. as the just created \verb+state+ object. Objects of the \verb+State+ class can be converted to a Python string and printed:
\begin{minted}{python}
print(state)
\end{minted}
\begin{verbatim}
    |1100>  *  1.0
    |1010>  *  1.0
    |1001>  *  1.0
    |0110>  *  1.0
    |0101>  *  1.0
    ...
\end{verbatim}
As for the \verb+Basis+ class, the number of shown determinants can be changed using the \\\verb+dets_printing_limit+ context manager. The underlying \verb+Basis+ object and the array of coefficients can be accessed directly as the attributes \verb+basis+ and \verb+coeffs+:
\begin{minted}{python}
print(state.basis)
\end{minted}
\begin{verbatim}
    1100
    1010
    1001
    0110
    0101
    ...
\end{verbatim}
\begin{minted}{python}
print(state.coeffs)
\end{minted}
\begin{verbatim}
    [1. 1. 1. 1. 1. 1.]
\end{verbatim}

Analogously to the \verb+Basis+ class demonstrated in the previous section, the \verb+State+ class implements the Python sequence protocol and supports NumPy-style ``fancy'' and boolean indexing~\cite{python_for_da_book}. From the quantum mechanical perspective, this functionality corresponds to projecting the state onto the Hilbert subspace spanned by the selected determinants.

Additionally, SOLAX supports operations of the form \verb+State % Basis+, which remove from the \verb+State+ object all determinants present in the \verb+Basis+ together with their associated coefficients. Quantum mechanically, this operation corresponds to projecting the state along the Hilbert subspace spanned by the deleted determinants. We demonstrate this operation using the \verb+basis_even+ and \verb+basis_odd+ objects from the previous section. The following code line retains in the resulting \verb+State+ only the determinants present in \verb+basis_odd+:
\begin{minted}{python}
state_odd = state % basis_even
print(state_odd)
\end{minted}
\begin{verbatim}
    |1010>  *  1.0
    |0110>  *  1.0
    |0011>  *  1.0
\end{verbatim}
\begin{minted}{python}
print(state_odd.basis == basis_odd)
\end{minted}
\begin{verbatim}
    True
\end{verbatim}
The equality operator \verb+==+ is not directly implemented for the \verb+State+ class because the latter involves real or complex coefficients that are represented up to machine precision. Treating these accurately is important to avoid unexpected behavior. Later in this section, we will demonstrate a method for comparing \verb+State+ objects with user-defined accuracy.

\subsubsection{Hilbert space operations}

The \verb+State+ class represents quantum many-body states and supports corresponding operations in the Hilbert space. 

\paragraph{A. Multiplication by a scalar.} A \verb+State+ object can be multiplied by a real or complex number resulting in a new \verb+State+ instance:
\begin{minted}{python}
print(2 * state)
\end{minted}
\begin{verbatim}
    |1100>  *  2.0
    |1010>  *  2.0
    |1001>  *  2.0
    |0110>  *  2.0
    |0101>  *  2.0
    ...
\end{verbatim}
The resulting \verb+State+ object contains a newly created NumPy array of \verb+coeffs+, while the underlying \verb+basis+ is shared between the new and original \verb+State+ objects. The scalar can be used as both the left and right operand. In addition to multiplication by a scalar, SOLAX also supports division by a scalar and the unary \verb+-+ operator.

\paragraph{B. Addition.} Two \verb+State+ objects can be added using the \verb|+| operator, which also applies the \verb|+| operator to their underlying \verb+Basis+ objects, as described in the previous section. The coefficients associated with the same Slater determinants in the operands are summed. For instance, consider two \verb+State+ objects obtained from \verb+state+ via ``fancy'' indexing:
\begin{minted}{python}
state1 = state[0, 1, 4]
print(state1)
\end{minted}
\begin{verbatim}
    |1100>  *  1.0
    |1010>  *  1.0
    |0101>  *  1.0
\end{verbatim}
\begin{minted}{python}
state2 = state[0, 3]
print(state2)
\end{minted}
\begin{verbatim}
    |1100>  *  1.0
    |0110>  *  1.0
\end{verbatim}
The addition operation gives:
\begin{minted}{python}
print(state1 + state2)
\end{minted}
\begin{verbatim}
    |1100>  *  2.0
    |1010>  *  1.0
    |0101>  *  1.0
    |0110>  *  1.0
\end{verbatim}
The subtraction operation based on addition and unary negation are also supported:
\begin{minted}{python}
print(state1 - state2)
\end{minted}
\begin{verbatim}
    |1100>  *  0.0
    |1010>  *  1.0
    |0101>  *  1.0
    |0110>  *  -1.0
\end{verbatim}

\paragraph{Important Note!} As demonstrated in the previous example, Slater determinants with zero coefficients are not automatically removed in SOLAX. Instead, determinants with exact zero coefficients, as well as those with very small coefficients, must be manually removed using a user-defined cutoff. This approach allows us to treat equivalently the following examples of "zeros" (where only the first is exactly zero as represented in the machine):
\begin{minted}{python}
print(0.1 + 0.1 - 0.2)
\end{minted}
\begin{verbatim}
    0.0
\end{verbatim}
\begin{minted}{python}
print(0.1 + 0.2 - 0.3)
\end{minted}
\begin{verbatim}
    5.551115123125783e-17
\end{verbatim}
We show in the following how to ``chop off'' such zeros using SOLAX tools.

\paragraph{C. Scalar product and normalization.} SOLAX supports Hermitian scalar product of \verb+State+ objects:
\begin{minted}{python}
print(state1 * state2)
\end{minted}
\begin{verbatim}
    1.0
\end{verbatim}
which allows to compute the norm as
\begin{minted}{python}
print(state * state)
\end{minted}
\begin{verbatim}
    6.0
\end{verbatim}
Once the norm is known, a \verb+State+ can be normalized using division by a scalar. SOLAX offers a shortcut for this operation implemented as the \verb+normalize+ method:
\begin{minted}{python}
state_normalized = state.normalize()
print(state_normalized * state_normalized)
\end{minted}
\begin{verbatim}
    1.0000000000000002
\end{verbatim}
The \verb+normalize+ method does not transform the initial \verb+State+ object but returns a new one. The underlying \verb+basis+ is shared.

\subsubsection{Chopping and equality of states}

Chopping a \verb+State+ object to a specified threshold is a useful operation implemented in SOLAX. As demonstrated below, the chop operation also enables the comparison of two \verb+State+ objects with a user-defined error margin.

\paragraph{A. The chop method.} The \verb+chop+ method removes all Slater determinants from a \verb+State+ object whose coefficients have absolute values smaller than a specified threshold. This operation creates a new \verb+State+ instance, leaving the original one unmodified. For demonstration, we consider the following \verb+State+:
\begin{minted}{python}
state3 = state2[-1]
state123 = state1 - state2 - state3
print(state123)
\end{minted}
\begin{verbatim}
    |1100>  *  0.0
    |1010>  *  1.0
    |0101>  *  1.0
    |0110>  *  -2.0
\end{verbatim}
Below we show 3 \verb+State+ objects obtained by chopping \verb+state123+ with respect to different thresholds:
\begin{minted}{python}
state_chopped1 = state123.chop(1e-14)
print(state_chopped1)
\end{minted}
\begin{verbatim}
    |1010>  *  1.0
    |0101>  *  1.0
    |0110>  *  -2.0
\end{verbatim}
\begin{minted}{python}
state_chopped2 = state123.chop(1.5)
print(state_chopped2)
\end{minted}
\begin{verbatim}
    |0110>  *  -2.0
\end{verbatim}
\begin{minted}{python}
state_chopped3 = state123.chop(2.5)
print(state_chopped3)
print(len(state_chopped3))
\end{minted}
\begin{verbatim}
    
    0
\end{verbatim}
As shown by the first example, the \verb+chop+ method allows the user to manually delete determinants with (numerically) zero coefficients. We stress again that in SOLAX this is not done automatically. 

\paragraph{B. Equality of two State objects.} As mentioned above, the direct equality relation \verb+==+ is not implemented for the \verb+State+ class due to the finite machine precision of the involved real or complex coefficients. Instead, we can construct the difference of the \verb+State+ objects and chop the result with respect to a small cutoff. If the obtained \verb+State+ is empty, then the compared \verb+State+ objects were close within the precision determined by the threshold. For demonstration, we use the following \verb+State+ objects \verb+state_a+ and \verb+state_b+ which are not exactly equal due to the machine error of the involved coefficients:
\begin{minted}{python}
state_mini = state[:2]
state_a = 0.1 * state_mini + 0.2 * state_mini
print(state_a)
\end{minted}
\begin{verbatim}
    |1100>  *  0.30000000000000004
    |1010>  *  0.30000000000000004
\end{verbatim}
\begin{minted}{python}
state_b = 0.3 * state_mini
print(state_b)
\end{minted}
\begin{verbatim}
    |1100>  *  0.3
    |1010>  *  0.3
\end{verbatim}
Following the described procedure, we obtain:
\begin{minted}{python}
state_diff = state_a - state_b
print(state_diff)
\end{minted}
\begin{verbatim}
    |1100>  *  5.551115123125783e-17
    |1010>  *  5.551115123125783e-17
\end{verbatim}
Now, chopping \verb+state_diff+ with respect to a small threshold gives an empty state:
\begin{minted}{python}
state_zero = state_diff.chop(1e-14)
print(len(state_zero))
\end{minted}
\begin{verbatim}
    0
\end{verbatim}
meaning that \verb+state_a+ and \verb+state_b+ are indeed equal within the error of \verb+1e-14+. Practice shows that proper control here can be crucial to avoid unexpected behavior.

\subsection{OperatorTerm}

We now introduce quantum mechanical operators represented in SOLAX by two classes: \\\verb+OperatorTerm+ and \verb+Operator+. The \verb+OperatorTerm+ class efficiently represents products of ladder operators with the same structure, while the \verb+Operator+ class encapsulates multiple \verb+OperatorTerm+ objects with different structures in a single entity. We will further discuss quantum operators as implemented in SOLAX through a concrete example, beginning with the \verb+OperatorTerm+ class.

Once more we consider the case of two electrons occupying four spin-orbitals. In this example, the four slots with indices 0, 1, 2, and 3 correspond to the electronic states $\uparrow^{(1)}$, $\downarrow^{(1)}$, $\uparrow^{(2)}$, and $\downarrow^{(2)}$, respectively. Here, the arrow and superscript indicate the spin and orbital quantum numbers, respectively. We examine the following hopping operator:
\begin{equation}
\hat{V} = v \left( \hat{a}^\dagger_0 \hat{a}_2 + \hat{a}^\dagger_1 \hat{a}_3 \right) + \mathrm{h. c.} \equiv v \left( \hat{a}^\dagger_{1\uparrow} \hat{a}_{2\uparrow} + \hat{a}^\dagger_{1\downarrow} \hat{a}_{2\downarrow} \right) + \mathrm{h. c.}
\end{equation}
and assume $v = 1$. For clarity, we additionally provided the formula in the format typically used in quantum mechanics via ``$\equiv$''. Note that $\hat{V}$ consists of ladder operator products of the same structure $\hat{a}_i^\dagger \hat{a}_j$ and, therefore, can be represented by one \verb+OperatorTerm+ object. We start with the non-Hermitian operator
\begin{equation}
\hat{V}_0 = \hat{a}^\dagger_0 \hat{a}_2 + \hat{a}^\dagger_1 \hat{a}_3\equiv \hat{a}^\dagger_{1\uparrow} \hat{a}_{2\uparrow} + \hat{a}^\dagger_{1\downarrow} \hat{a}_{2\downarrow} 
\end{equation}
and gradually build up $\hat{V}$ using the functionality provided in SOLAX.

\subsubsection{Object construction\label{sec:OperatorTerm_ObjConstruct}}

Here we show how \verb+OperatorTerm+ objects are instantiated by considering the example of the introduced $\hat{V}_0$ operator. However, to keep the explanation generic, we denote the number of multipliers in each ladder operator product as $L$, and the number of the summed products as $K$ (specifically for $\hat{V}_0$ we have $L=2$ and $K=2$).

Each \verb+OperatorTerm+ has 3 ingredients:
\begin{itemize}
 \item \verb+daggers+ is a tuple of 0s and 1s of length $L$ showing which ladder operators in the products are annihilation (0s) and which are creation (1s) operators;
 \item \verb+posits+ is a 2D NumPy array of shape $K \times L$ and integer type indicating at which positions (as counted from 0) the ladder operators in each product act;
 \item \verb+coeffs+ is a 1D NumPy array of length $K$ containing real or complex coefficients for each product.
\end{itemize}
For $\hat{V}_0$:
\begin{minted}{python}
daggers = (1, 0)

posits = np.array([
    [0, 2],
    [1, 3]
])

coeffs = np.array([
    1.0,
    1.0
])
\end{minted}
\noindent An \verb+OperatorTerm+ instance is then created as
\begin{minted}{python}
V0 = sx.OperatorTerm(daggers, posits, coeffs)
print(V0)
\end{minted}
\begin{verbatim}
    OperatorTerm(
        daggers=(1, 0),
        posits=array([[0, 2],
               [1, 3]]),
        coeffs=array([1., 1.])
    )
\end{verbatim}
We stress that in SOLAX, the ordering of ladder operators is fully the choice of the user which is controlled by the argument \verb+daggers+ indicating the positions of the creation and annihilation operators.

Unlike the \verb+Basis+ and \verb+State+ classes, which contain NumPy arrays with encoded Slater determinants that are not directly readable by the user, the \verb+OperatorTerm+ class displays its underlying NumPy arrays without any special formatting when converted to a Python string and printed. Additionally, if the \verb+posits+ array passed to \verb+OperatorTerm+ contains repeated rows, these duplicates are automatically removed, and the corresponding coefficients in \verb+coeffs+ are summed.

\subsubsection{Similarities with the State class}

The \verb+OperatorTerm+ class has strong similarities with the \verb+State+ class. Conceptually, they both represent expansions over some basis elements (ladder operator products and Slater determinants, respectively). At the technical level, both classes incorporate a 2D NumPy array with an accompanying 1D coefficient array. We list here the analogous features without going into comprehensive details.
\begin{itemize}
\item \verb+OperatorTerm+ implements the Python sequence protocol and supports the NumPy-style ``fancy'' and boolean indexing~\cite{python_for_da_book} (see also the section on the \verb+Basis+ class).
\item  Objects of \verb+OperatorTerm+ can be added using the \verb|+| operator. If the \verb+daggers+ tuples of the summands are equal (i.e. the quantum operators have the same structure), the result is of the \verb+OperatorTerm+ type and otherwise of the \verb+Operator+ type (the latter class is considered in the next section).
\item  The \verb+OperatorTerm+ class supports multiplication with scalars.
\item  \verb+OperatorTerm+ objects can be ``chopped'' with respect to a real threshold using the \verb+chop+ method.
\item Equality relation \verb+==+ is not implemented for the \verb+OperatorTerm+ class. As for the \verb+State+ class, the equality up to a user-determined precision can be checked with the binary \verb+-+ operator and subsequent chopping.
\end{itemize}

\subsubsection{Hermitian conjugate}

The \verb+OperatorTerm+ class supports the operation of Hermitian conjugation via the \verb+hconj+ property returning a new \verb+OperatorTerm+ object:
\begin{minted}{python}
print(V0.hconj)
\end{minted}
\begin{verbatim}
    OperatorTerm(
        daggers=(1, 0),
        posits=array([[2, 0],
               [3, 1]]),
        coeffs=array([1., 1.])
    )
\end{verbatim}
Using the introduced functionality, we can construct now the full operator $\hat{V}$ as
\begin{minted}{python}
V = V0 + V0.hconj
print(V)
\end{minted}
\begin{verbatim}
    OperatorTerm(
        daggers=(1, 0),
        posits=array([[0, 2],
               [1, 3],
               [2, 0],
               [3, 1]]),
        coeffs=array([1., 1., 1., 1.])
    )
\end{verbatim}
Note that the Hermitian conjugate $\hat{V}_0^\dagger$ has the same operator structure as $\hat{V}_0$, and hence the sum is represented by an \verb+OperatorTerm+ object.

\subsubsection{Acting on states and bases}

\verb+OperatorTerm+ objects represent quantum mechanical operators, meaning they can act on quantum states. To illustrate this, we consider the singlet state $|\Psi \rangle$, which is a normalized, anti-symmetric combination of a spin-up and spin-down electron, ensuring that the total spin $S_z = 0$:
\begin{minted}{python}
basis = sx.Basis(["1001", "0110"])
cfs = np.array([1.0, -1.0])

psi = sx.State(basis, cfs)
psi = psi.normalize()

print(psi)
\end{minted}
\begin{verbatim}
    |1001>  *  0.7071067811865475
    |0110>  *  -0.7071067811865475
\end{verbatim}
The operator action $\hat{V}| \Psi \rangle$ can be now performed as a call
\begin{minted}{python}
result_psi = V(psi)
print(result_psi)
\end{minted}
\begin{verbatim}
    |1100>  *  1.414213562373095
    |0011>  *  1.414213562373095
\end{verbatim}
Additionally, it is possible to act with an \verb+OperatorTerm+ directly on \verb+Basis+ objects. The result is then also of type \verb+Basis+ and contains the same Slater determinants as when acting on a \verb+State+:
\begin{minted}{python}
result_basis = V(psi.basis)
print(result_basis)
\end{minted}
\begin{verbatim}
    1100
    0011
\end{verbatim}
As will be demonstrated later in this work, the operation of acting directly on \verb+Basis+ objects is useful for iterative basis extension procedures via acting with operators.

\subsubsection{GPU acceleration and batches}

Acting with \verb+OperatorTerm+ is implemented using the JAX library~\cite{jax2018github} which supports computations on an NVIDIA GPU. Therefore, if such GPU is available on the machine and a GPU-capable version of JAX is installed, it will be automatically used for this operation.

\paragraph{Batching.} Since GPU memory is often scarce, we provide the user with the possibility to perform the operator action in batches by using the call arguments \verb+det_batch_size+ and \verb+op_batch_size+. They are responsible for batching the \verb+State+ (or \verb+Basis+) object and the \verb+OperatorTerm+ object, respectively. Note that these are keyword-only arguments and have to be provided with the argument names explicitly. For example:
\begin{minted}{python}
result_psi_batches = V(psi, det_batch_size=1, op_batch_size=2)
\end{minted}
\noindent We can now use the standard procedure to ensure that the \verb+State+ objects obtained with and without batching are equal within a very small error:
\begin{minted}{python}
s = result_psi_batches - result_psi
s = s.chop(1e-14)
print(len(s))
\end{minted}
\begin{verbatim}
    0
\end{verbatim}
Note, however, that the internal ordering of the Slater determinants in the resulting objects may be different as computed with and without batching. We note also that in this demonstration example, very small batch sizes are chosen, but in general, the user should aim at maximally exhausting the GPU memory.

\paragraph{Multiple GPUs.} Parallelization across multiple GPUs is currently under development and is made partially available in the current SOLAX version as an experimental feature. The action of an \verb+OperatorTerm+ can be enabled for multi-GPU mode by setting the keyword-only argument \verb+multiple_devices=True+. In this mode, the batches of the \verb+State+ (or \verb+Basis+) object are automatically distributed across all available GPUs on the machine. Note that for multi-GPU processing, the argument \verb+det_batch_size+ must be specified for creation of multiple batches. Otherwise, only one batch will be present and therefore only one GPU will be used. We stress again that this functionality is at the moment under development and is partially accessible as an experimental feature.

\subsection{Operator}

Thus far, we have considered quantum operators consisting of ladder operator products with the same structure, differing only in the positions on which the ladder operators act. These are represented in SOLAX using the \verb+OperatorTerm+ class. As the next step, we introduce the \verb+Operator+ class, which encapsulates \verb+OperatorTerm+ objects of different structures along with a scalar term. An \verb+Operator+ object can encode any quantum operator expressed in the second quantization formalism using annihilation and creation operators.

While computations can be performed using only the \verb+OperatorTerm+ class introduced in the previous section, we recommend the users to follow these guidelines:
\begin{itemize}
\item prefer the \verb+Operator+ class for basic usage like acting on quantum states or basis sets;
\item access the underlying \verb+OperatorTerm+ objects to fine-tune the operator.
\end{itemize}
In this section we will demonstrate this approach in practice.

\subsubsection{Construction of simple operators}

The basic way to instantiate \verb+Operator+ objects is based on the same ingredients as for the \verb+OperatorTerm+ class, i.e. \verb+daggers+, \verb+posits+ and \verb+coeffs+ (see Section~\ref{sec:OperatorTerm_ObjConstruct}). These arguments can be passed directly to the \verb+Operator+ constructor:
\begin{minted}{python}
op = sx.Operator(daggers, posits, coeffs)
print(op)
\end{minted}
\begin{verbatim}
    Operator({
        (1, 0): OperatorTerm(
            daggers=(1, 0),
            posits=array([[0, 2],
                   [1, 3]]),
            coeffs=array([1., 1.])
        )
    })
\end{verbatim}

\paragraph{Explanation of the printing output.} From the technical perspective, the \verb+Operator+ class implements the Python mapping protocol. In practice, this means that its objects behave similarly to Python dictionaries. Specifically, \verb+Operator+ objects store their underlying \verb+OperatorTerm+ objects as values in key-value pairs, with the keys being the corresponding \verb+daggers+ tuples. It is also possible to include a scalar term, which is associated with the string key \verb+"scalar"+. This structure is reflected when \verb+Operator+ objects are converted to a Python string and printed, as demonstrated in the example above.

Once \verb+daggers+, \verb+posits+ and \verb+coeffs+ are received, SOLAX creates automatically an \\\verb+OperatorTerm+ from the provided arguments and wraps it in an \verb+Operator+ object for convenient usage and further extension. In particular, this can be seen from the printed output for the \verb+op+ object above. Alternatively, the same \verb+Operator+ can be instantiated directly from the \verb+OperatorTerm+:
\begin{minted}{python}
print(sx.Operator(V0))
\end{minted}
\begin{verbatim}
Operator({
    (1, 0): OperatorTerm(
        daggers=(1, 0),
        posits=array([[0, 2],
               [1, 3]]),
        coeffs=array([1., 1.])
    )
})
\end{verbatim}
Here we reused the \verb+OperatorTerm+ object \verb+V0+ created in the previous section.

\subsubsection{Addition as a way to build operators}

The \verb+Operator+ created above is still trivial in the sense that it hosts only one \verb+OperatorTerm+. More advanced \verb+Operator+ objects can be constructed from more basic ones using addition. Also addition of an \verb+Operator+ with an \verb+OperatorTerm+ or a scalar leads to another \verb+Operator+. Moreover, as mentioned in the previous section, addition of two incompatible \verb+OperatorTerm+ objects does not lead to an error but creation of an \verb+Operator+ hosting the summands. For illustration, we introduce the following onsite energy operator:
\begin{equation}
\hat{U} = u_{01} \, \hat{a}^\dagger_0\hat{a}_0 \hat{a}^\dagger_1 \hat{a}_1 + u_{23} \, \hat{a}^\dagger_2 \hat{a}_2 \hat{a}^\dagger_3 \hat{a}_3 \equiv u_{1} \, \hat{a}^\dagger_{1\uparrow} \hat{a}_{1\uparrow}\hat{a}^\dagger_{1\downarrow} \hat{a}_{1\downarrow} + u_{2} \, \hat{a}^\dagger_{2\uparrow} \hat{a}_{2\uparrow}\hat{a}^\dagger_{2\downarrow} \hat{a}_{2\downarrow}
\end{equation}
with $u_{01} = 0.25 \equiv u_{1}$ and $u_{23} = 0.75 \equiv u_{2}$ as an example. As before, we additionally provided the formula in the format typically used in quantum mechanics via ``$\equiv$''. We encode $\hat{U}$ as an \verb+Operator+ object using the described standard way:
\begin{minted}{python}
daggers_u = (1, 0, 1, 0)
posits_u = np.array([
    [0, 0, 1, 1],
    [2, 2, 3, 3]
])
coeffs_u = np.array([0.25, 0.75])

U = sx.Operator(daggers_u, posits_u, coeffs_u)
print(U)
\end{minted}
\begin{verbatim}
    Operator({
        (1, 0, 1, 0): OperatorTerm(
            daggers=(1, 0, 1, 0),
            posits=array([[0, 0, 1, 1],
                   [2, 2, 3, 3]]),
            coeffs=array([0.25, 0.75])
        )
    })
\end{verbatim}
Now we can construct e.g. the compound operator
\begin{equation}
\hat{H} = \mathbb{1} + \hat{V} + \hat{U} = \mathbb{1} + \hat{V}_0 + \hat{V}^\dagger_0 + \hat{U}
\end{equation}
directly as
\begin{minted}{python}
H = 1 + V0 + V0.hconj + U
print(H)
\end{minted}
\begin{verbatim}
    Operator({
        (1, 0): OperatorTerm(
            daggers=(1, 0),
            posits=array([[0, 2],
                   [1, 3],
                   [2, 0],
                   [3, 1]]),
            coeffs=array([1., 1., 1., 1.])
        ),
        scalar: 1,
        (1, 0, 1, 0): OperatorTerm(
            daggers=(1, 0, 1, 0),
            posits=array([[0, 0, 1, 1],
                   [2, 2, 3, 3]]),
            coeffs=array([0.25, 0.75])
        )
    })
\end{verbatim}
As seen from the printed output, the resulting \verb+Operator+ consists of the scalar 1 and two \verb+OperatorTerm+ objects.

\subsubsection{Similarities with the OperatorTerm class}

The \verb+Operator+ class contains operations which are similar to the \verb+OperatorTerm+ class.
\begin{itemize}
\item Linear operations of addition and multiplication with scalars.
\item Hermitian conjugation using the \verb+hconj+ property.
\item Action on \verb+State+ and \verb+Basis+ objects --- is delegated to the underlying \verb+OperatorTerm+ components and multiplication with the scalar with subsequent addition of the partial results. Note that in case of action on a \verb+Basis+, the scalar, if present, acts effectively as the unity operator (even if it is equal to zero).
\item The keyword-only batching arguments \verb+det_batch_size+ and \verb+op_batch_size+ are available for the \verb+Operator+ action, and control batching for the \verb+OperatorTerm+ components.
\item We remind the user here, that the underlying \verb+OperatorTerm+ objects automatically support computations on an NVIDIA GPU.
\item For leveraging the multi-GPU parallelization (experimental!), the \verb+Operator+ action call can receive the keyword-only argument \verb+multiple_devices=True+ which propagates to the underlying \verb+OperatorTerm+ objects. In this case, batches corresponding to \\\verb+det_batch_size+ will be distributed over a few GPUs, if available.
\end{itemize}

\subsubsection{Manipulations with Operator objects}

\paragraph{A. Accessing underlying components.} A key difference between the \verb+Operator+ and \\\verb+OperatorTerm+ classes is the container type they implement. \verb+OperatorTerm+ objects are sequences and can be indexed using integer positions and slices, as well as NumPy's ``fancy'' and boolean indexing mechanisms. In contrast, the \verb+Operator+ class implements the Python mapping protocol, making it similar to Python dictionaries, which are indexed via their keys. For \verb+Operator+ objects, the keys are either the \verb+daggers+ tuples or the \verb+"scalar"+ string. For example:
\begin{minted}{python}
print(H[1, 0, 1, 0])
\end{minted}
\begin{verbatim}
    OperatorTerm(
        daggers=(1, 0, 1, 0),
        posits=array([[0, 0, 1, 1],
               [2, 2, 3, 3]]),
        coeffs=array([0.25, 0.75])
    )
\end{verbatim}
\begin{minted}{python}
print(H["scalar"])
\end{minted}
\begin{verbatim}
    1
\end{verbatim}
Note that for indexing, the ``tuple'' parenthesis for \verb+daggers+ can be omitted as in the example above. As usual Python dictionaries, \verb+Operator+ instances support views \verb+keys+, \verb+values+ and \verb+items+ for iteration over their entries.

\paragraph{B. Operator length.} \verb+Operator+ objects as mappings have length, which however only reflects the number of the stored components and is not related to the length of the underlying \verb+OperatorTerm+ objects. Indeed:
\begin{minted}{python}
print(len(H))
\end{minted}
\begin{verbatim}
    3
\end{verbatim}
whereas for the contained \verb+OperatorTerm+ components the lengths are:
\begin{minted}{python}
for key, term in H.items():
    if key != "scalar":
        print(f"Length of the OperatorTerm {key} is {len(term)}")
\end{minted}
\begin{verbatim}
    Length of the OperatorTerm (1, 0) is 4
    Length of the OperatorTerm (1, 0, 1, 0) is 2
\end{verbatim}

\paragraph{C. Dropping components.} We have seen how \verb+Operator+ instances can be enriched with new components using addition. Conversely, if there is a need to remove an \verb+OperatorTerm+ or the scalar, this can be done using the \verb+drop+ method, which takes the key of the component to be removed and returns a new \verb+Operator+ object without it. The original \verb+Operator+ remains unmodified. For example:
\begin{minted}{python}
H_without_V = H.drop(1, 0)
print((1, 0) in H_without_V)
\end{minted}
\begin{verbatim}
    False
\end{verbatim}
\begin{minted}{python}
H_without_scalar = H.drop("scalar")
print("scalar" in H_without_scalar)
\end{minted}
\begin{verbatim}
    False
\end{verbatim}
Here we used the Python \verb+in+ keyword for membership checks. This becomes automatically possible since the \verb+Operator+ class implements the mapping interface.

\paragraph{D. Chopping OperatorTerm components.} The chopping operation is implemented for the \verb+Operator+ class as the \verb+chop+ method, but can be applied only to a particular \verb+OperatorTerm+ via its \verb+daggers+ key. The returned \verb+Operator+ object contains the chopped version of the corresponding \verb+OperatorTerm+ or does not contain it at all if it has become empty after chopping. The initial \verb+Operator+ stays unchanged. Chopping for the scalar term is not supported. For instance, chopping \verb+U+ with respect to the cutoff 0.5 is performed as
\begin{minted}{python}
H_chopped1 = H.chop((1, 0, 1, 0), 0.5)
print(len(H_chopped1[1, 0, 1, 0]))
\end{minted}
\begin{verbatim}
    1
\end{verbatim}
leaving only one entry out of the two in the corresponding \verb+OperatorTerm+. Chopping with respect to 1.0 leads to chopping it off completely:
\begin{minted}{python}
H_chopped2 = H.chop((1, 0, 1, 0), 1.0)
print((1, 0, 1, 0) in H_chopped2)
\end{minted}
\begin{verbatim}
    False
\end{verbatim}

\subsection{OperatorMatrix}

To solve a quantum many-body eigenvalue problem, it is often necessary to construct the matrix representation of a quantum mechanical operator (e.g. Hamiltonian) on a given basis set. We address this requirement with the SOLAX class \verb+OperatorMatrix+, which provides tools for efficient matrix construction. It is important to note that the \verb+OperatorMatrix+ class is designed solely for the efficient construction of the operator matrix, while subsequent diagonalization is performed by the user with the help of the SciPy library~\cite{scipy}.

\subsubsection{Obtaining an OperatorMatrix}

For the demonstration, we reconsider again the case of two electrons in four spin-orbitals. We use the \verb+Operator+ object \verb+H+ constructed in the last section and create here also a \verb+Basis+ of Slater determinants with the spin projection $S_z = 0$:
\begin{minted}{python}
basis = sx.Basis(["1001", "1100", "0110", "0011"])
\end{minted}
\noindent The matrix of the operator $\hat{H}$ on this basis can be built directly using the \verb+build_matrix+ method of the corresponding \verb+Operator+ object (this method is available also for the \\\verb+OperatorTerm+ class):

\begin{minted}{python}
matrix = H.build_matrix(basis)
\end{minted}
\noindent The result is an object is of the \verb+OperatorMatrix+ class, which stores matrix elements in a coordinate sparse format, meaning only non-zero matrix elements are stored. Here, we highlight the main features of the matrix construction operation:

\begin{itemize}
\item It is possible to construct non-square matrices using two distinct \verb+Basis+ objects for rows and columns by passing them as arguments to the \verb+build_matrix+ method.
\item The \verb+build_matrix+ function supports the keyword-only arguments \verb+det_batch_size+ and \verb+op_batch_size+ which are used in evaluation of the matrix elements via action of the \verb+OperatorTerm+ objects (see the section on the \verb+OperatorTerm+ class).
\item Matrix evaluation inherits from the \verb+OperatorTerm+ class the possibility to perform computations automatically on an NVIDIA GPU, if available.
\item Computations on multiple GPUs (experimental!) can be switched on by providing \\\verb+multiple_devices=True+ to the \verb+build_matrix+ method. In this case, batches corresponding to \verb+det_batch_size+ will be distributed over a few GPUs, if available (see the section on the \verb+OperatorTerm+ class).
\end{itemize}
The dimensions of the constructed matrix can be accessed using the read-only \verb+size+ property:
\begin{minted}{python}
print(matrix.size)
\end{minted}
\begin{verbatim}
    (4, 4)
\end{verbatim}
The number of the non-zero matrix elements (i.e. the total number of the stored matrix elements) can be obtained as
\begin{minted}{python}
print(matrix.num_nonzero)
\end{minted}
\begin{verbatim}
    12
\end{verbatim}
The content of an \verb+OperatorMatrix+ object can be viewed after conversion to SciPy and NumPy which we discuss in the following.

\subsubsection{Conversion to SciPy and NumPy}

The built \verb+OperatorMatrix+ can be now converted to the SciPy format using the \verb+to_scipy+ method:
\begin{minted}{python}
coo_matrix = matrix.to_scipy()
\end{minted}
\noindent The returned object is of type \verb+scipy.sparse.coo_matrix+, and stores the matrix in the coordinate sparse format similarly to the \verb+OperatorMatrix+ class. If necessary, the user can now convert it to a different sparse format using the SciPy means. Here, we convert the matrix to the usual NumPy dense format and print it:
\begin{minted}{python}
dense_matrix = coo_matrix.todense()
print(dense_matrix)
\end{minted}
\begin{verbatim}
    [[ 1.    1.    0.    1.  ]
     [ 1.    1.25 -1.    0.  ]
     [ 0.   -1.    1.   -1.  ]
     [ 1.    0.   -1.    1.75]]
\end{verbatim}
Note that the NumPy dense representation is feasible only for small matrices, e.g. for demonstration or testing purposes. In this section we will often perform such conversion of \\\verb+OperatorMatrix+ objects in order to print their content in a usual matrix format. Therefore, we define the shortcut function:
\begin{minted}{python}
def print_matrix(m):
    print(m.to_scipy().todense())
\end{minted}

\subsubsection{Manipulations with OperatorMatrix objects}

Once obtained from an \verb+Operator+ or \verb+OperatorTerm+, the \verb+OperatorMatrix+ object allows for further useful manipulations, which we present here. In all examples considered below, a new transformed \verb+OperatorMatrix+ instance is created while the original remains unmodified. Here, we always utilize the \verb+matrix+ object constructed and shown above.

\paragraph{A. Displace.} The \verb+displace+ method allows to shift the matrix content along the row and the column axes with the corresponding change of the matrix dimensions. The two method arguments are the number of positions the matrix is displaced by vertically (row axis) and horizontally (column axis), respectively. The shifts can be positive and negative. For example:
\begin{minted}{python}
print_matrix(
    matrix.displace(2, 1)
)
\end{minted}
\begin{verbatim}
    [[ 0.    0.    0.    0.    0.  ]
     [ 0.    0.    0.    0.    0.  ]
     [ 0.    1.    1.    0.    1.  ]
     [ 0.    1.    1.25 -1.    0.  ]
     [ 0.    0.   -1.    1.   -1.  ]
     [ 0.    1.    0.   -1.    1.75]]
\end{verbatim}
\begin{minted}{python}
print_matrix(
    matrix.displace(-1, -1)
)
\end{minted}
\begin{verbatim}
    [[ 1.25 -1.    0.  ]
     [-1.    1.   -1.  ]
     [ 0.   -1.    1.75]]
\end{verbatim}
As seen from these examples, the newly created positions are effectively filled with zeros, whereas the entries with resulting negative positions are dropped.

\paragraph{B. Window.} The \verb+window+ method implements a rectangular filter, which sets all elements outside this rectangle to zero without changing the matrix shape. Technically, the filtered out matrix elements are directly discarded, since only non-zero matrix elements are stored in \verb+OperatorMatrix+ objects. For instance:
\begin{minted}{python}
print_matrix(
    matrix.window((1, 1), (3, 4))
)
\end{minted}
\begin{verbatim}
    [[ 0.    0.    0.    0.  ]
     [ 0.    1.25 -1.    0.  ]
     [ 0.   -1.    1.   -1.  ]
     [ 0.    0.    0.    0.  ]]
\end{verbatim}
The tuples passed to the \verb+window+ method correspond to the positions of the left upper (inclusive) and the right lower (exclusive) corners of the filter. If tuples $(a, b)$ and $(c, d)$ are passed, then the matrix values at the intersection of rows $i$: $a \le i < c$ and columns $j$: $b \le j < d$ survive, whereas the other matrix elements become zero. If any of the arguments $a$, $b$, $c$, $d$ is \verb+None+, it will be replaced by a position leading to the largest possible filter size.

\paragraph{C. Shrink basis.} After the matrix on a particular basis is constructed, it is straightforward to obtain the matrix on any sub-basis by extracting the corresponding matrix elements. This can be done using the \verb+shrink_basis+ method as we show in the following. We choose the sub-basis of our \verb+basis+ object corresponding to the electrons occupying different spatial orbitals:
\begin{minted}{python}
sub_basis = basis[0, 2]
print(sub_basis)
\end{minted}
\begin{verbatim}
    1001
    0110
\end{verbatim}
Then the matrix of the same operator $\hat{H}$ on this sub-basis can be obtained as
\begin{minted}{python}
print_matrix(
    matrix.shrink_basis(basis, sub_basis)
)
\end{minted}
\begin{verbatim}
    [[1. 0.]
     [0. 1.]]
\end{verbatim}
By default, this operation shrinks the basis along both axes. It is also possible to shrink the basis only along the row or the column axis by passing an additional argument \verb+axis=0+ or \verb+axis=1+ to the \verb+shrink_basis+ method, respectively.

\paragraph{Important note!} Both the sub-basis and the initial basis have to be passed to the \\\verb+shrink_basis+ method. Therefore, it is important to perform the sub-matrix extraction prior to any matrix displacements, since displaced matrices are not related to the initial basis anymore.

\paragraph{D. Chopping.} The \verb+chop+ method provides the possibility to discard matrix elements with absolute values less than a user-defined threshold. These entries are effectively set to zero (we remind that zeros are not stored in \verb+OperatorMatrix+ objects). As an example, we chop our matrix with respect to the threshold \verb+1.1+:
\begin{minted}{python}
print_matrix(
    matrix.chop(1.1)
)
\end{minted}
\begin{verbatim}
    [[0.   0.   0.   0.  ]
     [0.   1.25 0.   0.  ]
     [0.   0.   0.   0.  ]
     [0.   0.   0.   1.75]]
\end{verbatim}

\subsubsection{Linear operations and Hermitian conjugate}

Like other SOLAX classes related to quantum mechanical operators, the \verb+OperatorMatrix+ class supports addition, multiplication by real or complex scalars, and Hermitian conjugation via the \verb+hconj+ property. However, it is important to note that the addition operation deviates from the standard linear algebra convention. Specifically, \verb+OperatorMatrix+ objects can be added regardless of their shape. If the dimensions do not match, the matrices are effectively padded with zeros to form a minimal rectangle that encompasses both matrices before being added. This behavior is inherited from and natural for the \verb+OperatorMatrix+ implementation in the sparse format. To demonstrate this feature, we create two matrices of size $3 \times 3$ and $5 \times 2$ from our initial $4 \times 4$ matrix using displacements and then add them to obtain a matrix of size $5 \times 3$:

\begin{minted}{python}
matrix3_3 = matrix.displace(-1, -1)
print_matrix(matrix3_3)
\end{minted}
\begin{verbatim}
    [[ 1.25 -1.    0.  ]
     [-1.    1.   -1.  ]
     [ 0.   -1.    1.75]]
\end{verbatim}
\begin{minted}{python}
matrix5_2 = matrix.displace(1, -2)
print_matrix(matrix5_2)
\end{minted}
\begin{verbatim}
    [[ 0.    0.  ]
     [ 0.    1.  ]
     [-1.    0.  ]
     [ 1.   -1.  ]
     [-1.    1.75]]
\end{verbatim}
\begin{minted}{python}
print_matrix(
    matrix3_3 + matrix5_2
)
\end{minted}
\begin{verbatim}
    [[ 1.25 -1.    0.  ]
     [-1.    2.   -1.  ]
     [-1.   -1.    1.75]
     [ 1.   -1.    0.  ]
     [-1.    1.75  0.  ]]
\end{verbatim}
Later in this work we will use the introduced manipulations for efficient computations with operator matrices.

\subsubsection{Equality of OperatorMatrix objects}

As also the other SOLAX classes containing real or complex numbers, the \verb+OperatorMatrix+ class does not directly support the equality relation and the \verb+==+ operator. The user can check closeness of non-zero elements in two \verb+OperatorMatrix+ objects by

\begin{enumerate}[a)]
\item subtracting them;
\item using the \verb+chop+ method with respect to a small threshold;
\item ensuring that the \verb+num_nonzeros+ property returns zero.
\end{enumerate}
Note that due to the specific \verb+OperatorMatrix+ implementation, this procedure disregards completely the matrix dimensions and only checks closeness of non-zero elements having the same position in the matrices. For example, consider the following matrices \verb+m1+ and \verb+m2+:
\begin{minted}{python}
m1 = matrix.window((0, 0), (2, 2))
print_matrix(m1)
\end{minted}
\begin{verbatim}
    [[1.   1.   0.   0.  ]
     [1.   1.25 0.   0.  ]
     [0.   0.   0.   0.  ]
     [0.   0.   0.   0.  ]]
\end{verbatim}
\begin{minted}{python}
m2 = matrix.shrink_basis(basis, basis[:2])
print_matrix(m2)
\end{minted}
\begin{verbatim}
    [[1.   1.  ]
     [1.   1.25]]
\end{verbatim}
The outlined procedure gives
\begin{minted}{python}
print(
    (m1 - m2).chop(1e-14).num_nonzero == 0
)
\end{minted}
\begin{verbatim}
    True
\end{verbatim}
That is, all non-zero matrix elements having the same position are equal (within the chosen accuracy), but still the dimensions may differ, as in the considered case. The latter can be additionally compared as
\begin{minted}{python}
print(m1.size == m2.size)
\end{minted}
\begin{verbatim}
    False
\end{verbatim}

\subsection{Demonstration Computation for SIAM\label{sec:siam_noml}}

After the introduction of the foundational components of the SOLAX package, we now demonstrate its application to the Single Impurity Anderson Model (SIAM). This example illustrates how SOLAX can be used to construct and analyze finite size quantum systems by efficiently handling basis sets, states, and operators, specifically to find the ground state of a complex quantum many-body model.

\subsubsection{Introduction to SIAM}
The SIAM is a cornerstone model in the study of strongly correlated electron systems. Initially introduced by Anderson~\cite{Anderson1961}, the model describes a single localized level (the ``impurity'') with onsite interaction $U$ coupled to a continuum of noninteracting conduction electrons (the ``bath''). 
It was proposed to capture essential physics relevant to magnetic impurities in metals, such as those found in dilute alloys like gold doped with iron. Notably, it describes the Kondo effect, a phenomenon in which the impurity spin is screened by the surrounding conduction electrons at low temperatures, resulting in a highly correlated many-body ground state.
In addition to being a paradigm in its own right, the SIAM also serves as an auxiliary model for the dynamical mean-field (DMFT) solution of the Hubbard and related models \cite{dmft1,dmft2,dmft3}. 

Today, the most frequently used SIAM solvers are Quantum Monte Carlo methods which usually work on the imaginary Matsubara time domain, see e.g. Refs.~\cite{HFQMC,CTQMC}. While they allow for continuous baths, they often struggle with the fermionic sign problem at low temperatures \cite{SIGNPROBLEM}. Complementary diagonalization solvers like our approach do not suffer from any sign problem but require a discretized bath, which is represented by a finite number of single particle energy levels, each coupled to the impurity through hybridization terms. This introduces a computational challenge: the number of bath sites $N_\text{bath}$ controls the resolution of the bath representation, and pushing $N_\text{bath}$ to larger values is crucial to accurately capture the continuous nature of the bath, especially at low temperatures. Our SOLAX package is well-suited to this task, as it can efficiently handle large fermionic clusters. In turn, the possibility of incrementally enlarging the discretized SIAM model makes it an ideal and flexible test case for our code.

\subsubsection{Model Description}

\begin{figure}[!ht]
     \begin{center}
     \includegraphics[width=0.3\textwidth]{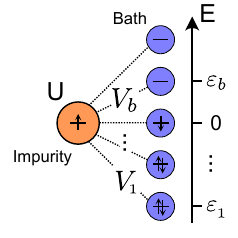}
     \end{center}
     \caption{Schematic illustration of SIAM with $N_\text{bath} = 5$ bath sites. The model is considered in the ``star geometry'', in which the correlated impurity (orange circle) hybridizes with the non-interacting and uncorrelated bath sites (blue circles). The impurity onsite energy is set to zero. The onsite impurity interaction is given by $U$. The bath site energies are $\varepsilon_b$ and the hybridization strengths are $V_b$.}
     \label{fig:siam}
\end{figure}

In Fig.~\ref{fig:siam} we sketch the SIAM which consists of an impurity with an effective onsite interaction $U$, coupled to a set of $N_\text{bath}$ non-interacting bath sites with energies $\varepsilon_b$ and hybridization strengths $V_b$. The SIAM Hamiltonian reads
\begin{eqnarray}
    \hat{H}_\text{SIAM}&=&\hat{H}_\mathrm{imp} + \hat{H}_\mathrm{bath} + \hat{H}_\mathrm{hyb}\;,\label{eq:h_full}\\
    \hat{H}_\mathrm{imp}&=& U\;\left(\hat{n}_{\text{imp}\uparrow}-\frac{1}{2}\right)\left(\hat{n}_{\text{imp}\downarrow}-\frac{1}{2}\right)\;,\label{eq:h_imp}\\
    \hat{H}_\mathrm{bath}&=& \sum_{\sigma\in\{\uparrow,\downarrow\}}\sum_{b=1}^{N_\text{bath}}  \varepsilon_b \hat{n}_{b\sigma}\;,\label{eq:h_bath}\\
    \hat{H}_\mathrm{hyb}&=& \sum_{\sigma\in\{\uparrow,\downarrow\}}\sum_{b=1}^{N_\text{bath}}V_{b}\left(\hat{a}_{\text{imp}\,\sigma}^\dagger \hat{a}_{b\sigma} + \text{h.c.}\right)\;.\label{eq:h_hyb}
\end{eqnarray}
where $\hat{a}^\dagger_{\alpha\sigma}$ and $\hat{a}_{\alpha\sigma}$ are fermionic creation and annihilation operators with $\alpha$ labeling the respective (impurity or bath) site, and $\hat{n}_{\alpha\sigma}\equiv \hat{a}^\dagger_{\alpha\sigma} \hat{a}_{\alpha\sigma}$ are the corresponding occupation operators. Note that we assumed here the onsite impurity energy to be zero. The parameters of the model are the number of non-interacting bath sites $N_\text{bath}$, the onsite energies of the bath sites $\varepsilon_\text{b}$, the hybridization amplitudes $V_b$,  and the particle-hole symmetric interaction on the impurity site $U$.

We follow Ref.~\cite{Bilous2024_SIAM} and choose the parameters $\varepsilon_b$ and $V_b$ such that our star geometry maps directly to a 1D chain i.e., an impurity site coupled to the first site of a 1D bath chain with hybridization $V$ with constant nearest neighbor hopping $t$ \cite{Aichhorn2011}. To this end, we set 
\begin{equation}
\begin{split}
    \varepsilon_b &= -2 t \cos{\left(\frac{b\pi}{N_\mathrm{bath}+1}\right)}\, ,
    \\
    V_b &= V\sqrt{\frac{2}{N_\mathrm{bath}+1}}\sqrt{1-\left(\frac{\varepsilon_b}{2t}\right)^2}\, ,
\end{split}
\end{equation}
with the bath site index $b$ running from 1 to $N_\text{bath}$. In the present work, we choose $V=\sqrt{10}$ eV and $t=1.0$ eV, and give all energies in units of $t$. Moreover, we restrict our calculations to an odd number of bath sites (such that there is always a bath site at $\varepsilon_{b}=0$) and half-filling, i.e. $N_{e}=N_\text{bath}+1$.
%
%

\begin{figure}[!ht]
     \begin{center}
     \includegraphics[width=0.6\textwidth]{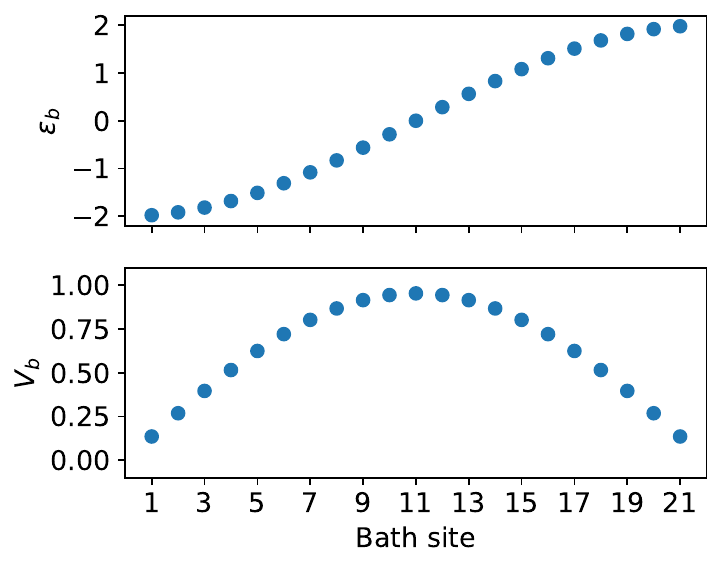}
     \end{center}
     \caption{The bath site energies $\varepsilon_b$ and the hybridization strengths $V_b$ for $N_\text{bath} = 21$.}
     \label{fig:bath}
\end{figure}

To set up the parameters for our bath we use the \verb+build_bath+ function:
\begin{minted}{python}
def build_bath(N_bath):
    ii = np.arange(N_bath) + 1
    xx = ii * np.pi / (N_bath + 1)
    e_bath = -2 * np.cos(xx)
    
    V0 = np.sqrt(20 / (N_bath + 1))
    V_bath = V0 * np.sqrt(1 - (e_bath / 2)**2)
    
    return e_bath, V_bath
\end{minted}
\noindent In Fig.~\ref{fig:bath} we show $\varepsilon_b$ and $V_b$ for the case of $N_\mathrm{bath} = 21$ as constructed using this function.

\subsubsection{Eigenvalue problem and solution procedure}

We aim to compute the energy of the ground state, which is known to belong to the $S_z = 0$ sector. An exact solution would require the construction and partial diagonalization of the Hamiltonian matrix over the basis set of all possible Slater determinants with $S_z = 0$. However, even with a few tens of bath sites, the complete basis becomes combinatorially large and numerically intractable. Therefore, it is common to perform these computations iteratively on a growing partial basis. Starting with an initial set of Slater determinants, the following iterations are performed:
\begin{itemize}
\item  construction and partial diagonalization of the Hamiltonian matrix on the current basis set;
\item extension of the basis set by acting on it with an extension operator $\hat{O}$ (here we choose $\hat{O} = \hat{H}$).
\end{itemize}
The energies obtained in each iteration at the diagonalization stage are monitored in order to stop the computation once convergence is achieved.

We mention in passing, that the choice of the initial set of Slater determinants, as well as the choice of the extension operator, can affect the convergence of the procedure \cite{Bilous2024_SIAM}. A natural choice for the initial set is the mean-field solution which consists of a single or a few degenerate Slater determinants. Below, we chose the twofold spin degenerate $S_z=0$ solution of the zero-hybridization limit (in the particle-hole symmetric case considered here) as our initial set of Slater determinants.

\subsubsection{Representation of Slater determinants\label{sec:SIAM_starting_SD}}

In order to represent Slater determinants as strings of 0s and 1s, we group together the spin-orbital pairs corresponding to the same orbital but having the opposite spins $\uparrow\downarrow$. The leftmost pair in the string is attributed to the impurity and the further pairs correspond to the bath sites with growing energies from left to right. The content of a determinantal string is illustrated as
\begin{equation}
\underbrace{\uparrow\downarrow}_\mathrm{imp.}\;\;\underbrace{\overbrace{\uparrow\downarrow \ldots \uparrow\downarrow}^{\varepsilon_b < 0}\;\;\overbrace{\uparrow\downarrow}^{\varepsilon_b = 0}\;\;\overbrace{\uparrow\downarrow \ldots \uparrow\downarrow}^{\varepsilon_b > 0}}_\mathrm{bath}
\end{equation}
As mentioned above, we take the two degenerate determinants with $S_z=0$ and the lowest net one-particle energy in the zero-hybridization limit as our initial basis:
\begin{equation}\label{eq:start_det1}
10\;\;11 \ldots 11\;\;01\;\;00 \ldots 00\;,
\end{equation}
\begin{equation}\label{eq:start_det2}
01\;\;11 \ldots 11\;\;10\;\;00 \ldots 00\;.
\end{equation}
For given $N_\mathrm{bath}$, Python strings for these determinants can be build using the function
\begin{minted}{python}
def build_start_dets(N_bath):
    det1 = "01"  + "1" * (N_bath - 1) + "10" + "0" * (N_bath - 1)
    det2 = "10"  + "1" * (N_bath - 1) + "01" + "0" * (N_bath - 1)
    return det1, det2
\end{minted}

\subsubsection{Starting Basis object}

The impurity onsite interaction strength is in all examples $U=10$. For the start, we stick to the simplest non-degenerate case of $N_\mathrm{bath} = 3$ bath sites and increase later $N_\mathrm{bath}$ for more advanced demonstrations.
\begin{minted}{python}
U = 10
N_bath = 3
e_bath, V_bath = build_bath(N_bath)
start_dets = build_start_dets(N_bath)
\end{minted}
\noindent We follow the standard procedures described in the previous sections to construct necessary objects of the SOLAX classes. In particular, a \verb+Basis+ object containing the two starting Slater determinants (\ref{eq:start_det1}, \ref{eq:start_det2}) is created as
\begin{minted}{python}
basis_start = sx.Basis(start_dets)
print(basis_start)
\end{minted}
\begin{verbatim}
    01111000
    10110100
\end{verbatim}

\subsubsection{Operator object for Hamiltonian}

Here we encode the Hamiltonian in parts represented by Eqs.~(\ref{eq:h_full})---(\ref{eq:h_hyb}). Note that for demonstration purposes, we switch the NumPy module to the regime in which up to 3 digits after the decimal point are printed (this does not influence the computational precision).

\paragraph{Impurity term.} We represent the impurity onsite interaction operator as 
\begin{equation}
\hat{H}_\mathrm{imp} =
    \underbrace{
        U\,\hat{a}^\dagger_{\mathrm{imp}{\uparrow}}\,\hat{a}_{\mathrm{imp}{\uparrow}}
        \hat{a}^\dagger_{\mathrm{imp}{\downarrow}}\,\hat{a}_{\mathrm{imp}{\downarrow}}
    }_{\hat{H}_\mathrm{imp}^{(2)}}
    -
    \underbrace{
        \frac{U}{2}
        \left(
        \hat{a}^\dagger_{\mathrm{imp}{\uparrow}}\,\hat{a}_{\mathrm{imp}{\uparrow}}
        +
        \hat{a}^\dagger_{\mathrm{imp}{\downarrow}}\,\hat{a}_{\mathrm{imp}{\downarrow}}
        \right)
    }_{\hat{H}_\mathrm{imp}^{(1)}}
    +
    \frac{U}{4}
\end{equation}
by expanding Eq.~(\ref{eq:h_imp}) and build up our \verb+Operator+ object term by term:
\begin{minted}{python}
H_imp2 = sx.Operator(
    (1, 0, 1, 0),
    np.array([
        [0, 0, 1, 1]
    ]),
    np.array([U])
)

H_imp1 = sx.Operator(
    (1, 0),
    np.array([
        [0, 0],
        [1, 1]
    ]),
    np.array([-U / 2, -U / 2])
)
\end{minted}
\begin{minted}{python}
H_imp = H_imp2 + H_imp1 + U / 4
print(H_imp)
\end{minted}
\begin{verbatim}
Operator({
    (1, 0): OperatorTerm(
        daggers=(1, 0),
        posits=array([[0, 0],
               [1, 1]]),
        coeffs=array([-5., -5.])
    ),
    (1, 0, 1, 0): OperatorTerm(
        daggers=(1, 0, 1, 0),
        posits=array([[0, 0, 1, 1]]),
        coeffs=array([10.])
    ),
    scalar: 2.5
})
\end{verbatim}
\paragraph{Bath term.} Taking into account that in each bath site both spin-orbitals have the same energy, the bath Hamiltonian term $\hat{H}_\mathrm{bath}$ is encoded as
\begin{minted}{python}
H_bath = sx.Operator(
    (1, 0),
    np.arange(2, 2 * N_bath + 2).repeat(2).reshape(-1, 2),
    e_bath.repeat(2)
)
print(H_bath)
\end{minted}
\begin{verbatim}
    Operator({
        (1, 0): OperatorTerm(
            daggers=(1, 0),
            posits=array([[2, 2],
                   [3, 3],
                   [4, 4],
                   [5, 5],
                   [6, 6],
                   [7, 7]]),
            coeffs=array([-1.414e+0, -1.414e+00, -1.225e-16, -1.225e-16,
                    1.414e+00, 1.414e+00])
        )
    })
\end{verbatim}

\paragraph{Hybridization term.} Hybridization as described by the $\hat{H}_\mathrm{hyb}$ term takes place between spin-orbitals with the same spin. An \verb+Operator+ object for $\hat{H}_\mathrm{hyb}$ without h.c. is then built as
\begin{minted}{python}
H_hyb_posits = np.vstack([
    np.array([0, 1] * N_bath),
    np.arange(2, 2 * N_bath + 2)
]).T

H_hyb_nohc = sx.Operator(
    (1, 0),
    H_hyb_posits,
    V_bath.repeat(2)
)
print(H_hyb_nohc)
\end{minted}
\begin{verbatim}
    Operator({
        (1, 0): OperatorTerm(
            daggers=(1, 0),
            posits=array([[0, 2],
                   [1, 3],
                   [0, 4],
                   [1, 5],
                   [0, 6],
                   [1, 7]]),
            coeffs=array([1.581, 1.581, 2.236, 2.236, 1.581, 1.581])
        )
    })
\end{verbatim}

\paragraph{Full Hamiltonian. } Finally we obtain an \verb+Operator+ object for the full SIAM Hamiltonian (which we don't print here):
\begin{minted}{python}
H = H_imp + H_bath + H_hyb_nohc + H_hyb_nohc.hconj
\end{minted}

\subsubsection{Hamiltonian matrix and state energy}

We can now obtain the Hamiltonian matrix on the basis of the 2 ``starting'' Slater determinants as
\begin{minted}{python}
matrix_start = H.build_matrix(basis_start)
\end{minted}
\noindent For this demonstration example, the obtained \verb+OperatorMatrix+ object can be converted to the NumPy dense format and printed:
\begin{minted}{python}
matrix_dense_start = matrix_start.to_scipy().todense()
print(matrix_dense_start)
\end{minted}
\begin{verbatim}
    [[-5.328  0.   ]
     [ 0.    -5.328]]
\end{verbatim}
This matrix is diagonal, and contains directly the state energy corresponding to the Hartree-Fock approximation:
\begin{minted}{python}
energy_start = matrix_dense_start[0, 0]
print(energy_start)
\end{minted}
\begin{verbatim}
    -5.32842712474619
\end{verbatim}

\subsubsection{Basis extension}

The state energy obtained above is the roughest approximation and must be refined by extending the basis to span a larger subspace of the Hilbert space of the considered many-body system. This can be achieved by generating new determinants via acting with an extension operator on the initial basis. Usually, extension operators are chosen which promote electrons from occupied to unoccupied orbitals via single or double excitation, see e.g. Ref.~\cite{Helgaker2000}. In our example, we use the Hamilton operator itself (which in this case includes only single excitations). The advantage of this choice is that the excitation procedure automatically respects the symmetry of the problem. Specifically, extension with the spin-conserving Hamiltonian $\hat{H}_\text{SIAM}$ from the two initial determinants with the total spin projection $S_z = 0$, yields only determinants with $S_z = 0$. In this way, we automatically generate only determinants with non-zero contribution to the ground state.
%

To this end, we extend the basis by acting on \verb+basis_start+ with the extension operator (i.e. the Hamiltonian) as
\begin{minted}{python}
basis = H(basis_start)
print(len(basis))
\end{minted}
\begin{verbatim}
    8
\end{verbatim}
The matrix built on this basis is not diagonal anymore:
\begin{minted}{python}
matrix = H.build_matrix(basis)

matrix_dense = matrix.to_scipy().todense()
print(matrix_dense)
\end{minted}
\begin{verbatim}
    [[-5.328 -1.581 -2.236 -2.236 -1.581  0.     0.     0.   ]
     [-1.581  1.086  0.     0.     0.     0.     0.     0.   ]
     [-2.236  0.    -0.328  0.     0.     2.236  0.     0.   ]
     [-2.236  0.     0.    -0.328  0.     2.236  0.     0.   ]
     [-1.581  0.     0.     0.     1.086  0.     0.     0.   ]
     [ 0.     0.     2.236  2.236  0.    -5.328 -1.581 -1.581]
     [ 0.     0.     0.     0.     0.    -1.581  1.086  0.   ]
     [ 0.     0.     0.     0.     0.    -1.581  0.     1.086]]
\end{verbatim}
Therefore, in order to obtain the state energy, the lowest eigenvalue has to be computed using the SciPy means (we use NumPy in this demonstration example):
\begin{minted}{python}
energy = np.linalg.eigvals(matrix_dense).min()

basis_size = len(basis)
print(f"Basis size = {basis_size}\tEnergy = {energy}")
\end{minted}
\begin{verbatim}
    Basis size = 8  Energy = -8.351171437060554
\end{verbatim}
The iterations of basis extension and Hamiltonian matrix evaluation should be now repeated until the state energy converges. We switch now to a more advanced example with larger $N_\text{bath}$ for demonstration of these iterations.

\subsubsection{An example of full computation\label{sec:siam_noml_full_comp}}

We now consider the SIAM with $N_\mathrm{bath} = 21$ bath sites and iteratively evaluate the state energy using the described approach. The reconstruction of the \verb+Basis+ and \verb+Operator+ objects can be performed directly by rerunning the code above after assigning the new value to the \verb+N_bath+ variable. We omit this part and show the loop with the iterations directly.
\begin{minted}{python}
import scipy as sp

num_iterations = 4

basis = basis_start

for i in range(num_iterations):
    matrix = H.build_matrix(basis)
    energy = sp.sparse.linalg.eigsh(
        matrix.to_scipy(), k=1, which="SA"
    )[0][0]
    
    basis_size = len(basis)
    print(
        f"Iteration: {i+1:<8d}"
        f"Basis size = {basis_size:<12d}"
        f"Energy = {energy}"
    )
    
    if i < num_iterations - 1:
        basis = H(basis)
\end{minted}
\begin{verbatim}
Iteration: 1       Basis size = 2           Energy = -28.463653910211487
Iteration: 2       Basis size = 44          Energy = -30.19530217404953
Iteration: 3       Basis size = 684         Energy = -31.242891311317756
Iteration: 4       Basis size = 7084        Energy = -31.70729257122757
\end{verbatim}
To find the ground state in each iteration, we used the SciPy diagonalization routine \\\verb+sp.sparse.linalg.eigsh+ for Hermitian sparse matrices.
We pass the Hamiltonian matrix and request the first smallest (\verb+k=1+, \verb+which="SA"+) eigenvalue. We note that the basis is not extended in the last iteration since the computation terminates immediately thereafter. It is seen that the energy is converging with the iterations, which should be stopped once the desired precision is achieved.

\subsubsection{Optimization of matrix construction\label{sec:siam_noml_matrix_optim}}

In the iterations above we rebuilt the Hamiltonian matrix each time. Using the \verb+OperatorMatrix+ tools demonstrated in the previous section, it is possible to avoid re-evaluation of the matrix elements which have already been calculated in the previous iteration.

To demonstrate this, we start from the \verb+OperatorMatrix+ object constructed in the last performed iteration. In Python, variables remain available after the loop is finished. Therefore, we access \verb+basis+ and \verb+matrix+ directly, and use different variable names corresponding to the analytical notations below:
\begin{minted}{python}
basis_small = basis
M_small = matrix
print(M_small.size)
\end{minted}
\begin{verbatim}
    (7084, 7084)
\end{verbatim}
In the following, we will use the analytical notation $M_\mathrm{small}$ corresponding to the \verb+OperatorMatrix+ object \verb+M_small+. Now we extend the basis another time and build the Hamiltonian matrix $M_\mathrm{big}^\mathrm{direct}$ directly on the resulting basis (thus computing all matrix elements from scratch as before):
\begin{minted}{python}
basis_big = H(basis_small)
M_big_direct = H.build_matrix(basis_big)
print(M_big_direct.size)
\end{minted}
\begin{verbatim}
    (58984, 58984)
\end{verbatim}
However, the set of Slater determinants \verb+basis_small+ is a subset of \verb+basis_big+; this can be checked e.g. as follows:
\begin{minted}{python}
print(len(basis_small % basis_big) == 0)
\end{minted}
\begin{verbatim}
    True
\end{verbatim}
Therefore, all matrix elements of $M_\mathrm{small}$ enter also $M_\mathrm{big}^\mathrm{direct}$ allowing to avoid unnecessary re-computations. We implement here a possible scenario of such optimized matrix construction. Note, however, that $M_\mathrm{small}$ is in general not a rectangular submatrix in $M_\mathrm{big}^\mathrm{direct}$. Instead, the matrix elements of $M_\mathrm{small}$ are spread in $M_\mathrm{big}^\mathrm{direct}$ according to the positions of the Slater determinants from \verb+basis_small+ in \verb+basis_big+. In the following we construct a matrix $M_\mathrm{big}$ which does contain $M_\mathrm{small}$ as a true submatrix. Though $M_\mathrm{big}$ and $M_\mathrm{big}^\mathrm{direct}$ are in general not exactly equal, they are equivalent up to permutation of the basis determinants irrelevant for our applications.

\paragraph{Evaluation of the missing submatrix.} The target matrix $M_\mathrm{big}$ is a Hermitian block matrix
\begin{equation}
M_\mathrm{big} = \begin{pmatrix}
M_\mathrm{small} & A\\
A^\dagger & B
\end{pmatrix}
\end{equation}
with unknown blocks $A$ and $B$, where $B$ is Hermitian. Given $M_\mathrm{small}$ is known, we need to additionally evaluate only the block matrix
\begin{equation}
C = \begin{pmatrix}
A\\
B
\end{pmatrix}
\end{equation}
in order to construct $M_\mathrm{big}$. The rectangular matrix $C$ is built on the following \verb+Basis+ objects:
\begin{minted}{python}
basis_cols = basis_big % basis_small
basis_rows = basis_small + basis_cols
\end{minted}
\noindent As mentioned in the section on the \verb+OperatorMatrix+ class, this can be achieved by passing both \verb+Basis+ objects to the \verb+build_matrix+ method:
\begin{minted}{python}
C = H.build_matrix(basis_rows, basis_cols)
print(C.size)
\end{minted}
\begin{verbatim}
    (58984, 51900)
\end{verbatim}

\paragraph{Constructing the matrix from its parts.} We use now the methods supported by the \\\verb+OperatorMatrix+ class to build up $M_\mathrm{big}$ from $M_\mathrm{small}$ and $C$. First of all, we bring \verb+C+ to its right place in $M_\mathrm{big}$ by displacing it along the column axis, and obtain the matrix
\begin{equation}
C_\mathrm{displ} = 
\begin{pmatrix}
0 & A\\
0 & B
\end{pmatrix}
\end{equation}
\begin{minted}{python}
C_displ = C.displace(0, len(basis_small))
\end{minted}
\noindent Then the following sum corresponds to the block matrix
\begin{equation}
M_\mathrm{with2B} = 
\begin{pmatrix}
M_\mathrm{small} & A\\
A^\dagger & 2B
\end{pmatrix}
\end{equation}
\begin{minted}{python}
M_with2B = M_small + C_displ + C_displ.hconj
\end{minted}
\noindent Now we need to subtract $B$ displaced to the proper position:
\begin{equation}
B_\mathrm{displ} = 
\begin{pmatrix}
0 & 0\\
0 & B
\end{pmatrix}
\end{equation}
This matrix can be obtained directly from $C_\mathrm{displ}$ using the \verb+window+ method:
\begin{minted}{python}
left_top = (len(basis_small), len(basis_small))
right_bottom = (None, None)

B_displ = C_displ.window(left_top, right_bottom)
\end{minted}
\noindent Then finally the targeted $M_\mathrm{big}$ matrix is obtained as
\begin{minted}{python}
M_big = M_with2B - B_displ
\end{minted}
\paragraph{Comparison of the results.} We ensure now the equality of the state energies obtained using the direct and the optimized approaches to the matrix construction:
\begin{minted}{python}
energy_big_direct = sp.sparse.linalg.eigsh(
    M_big_direct.to_scipy(), k=1, which="SA"
)[0][0]
print(energy_big_direct)
\end{minted}
\begin{verbatim}
    -31.819018639483936
\end{verbatim}
\begin{minted}{python}
energy_big = sp.sparse.linalg.eigsh(
    M_big.to_scipy(), k=1, which="SA"
)[0][0]
print(energy_big)
\end{minted}
\begin{verbatim}
    -31.819018639483833
\end{verbatim}
It is evident that the obtained results agree within a very small error, which can be attributed to randomization in the SciPy eigensolver.

In applications involving particularly large Hermitian operators, such as the Hamiltonian in the computations for the N$_2$ molecule performed in Ref.~\cite{Schmerwitz2024_N2}, working directly with full operators is disadvantageous. Instead, it is possible to represent a Hermitian operator $\hat{A}$ as the sum $\hat{A} = \hat{B} + \hat{B}^\dagger$ and use only the part $\hat{B}$ for computations. In particular, this greatly simplifies the construction of the operator matrix. Once an \verb+OperatorMatrix+ is built for $\hat{B}$, adding its Hermitian conjugate creates an \verb+OperatorMatrix+ for the full operator $\hat{A}$.

\section{Neural network support for tackling big basis sets\label{sec:nn_support}}

We have presented the basic functionality of SOLAX as a solver for fermionic quantum systems and now switch to the built-in neural network (NN) support for managing large sets of Slater determinants. While the iterative solution procedure demonstrated for the SIAM in the previous section allows energy refinement to arbitrary precision, the practical implementation of the basis extension approach leads to exponential growth in the basis size, making computations increasingly infeasible. In this section, we demonstrate the NN-based tools available in SOLAX to control basis growth and converge the results with reduced computational resources.

To this end, we follow the algorithm developed in Ref.~\cite{Bilous2024_SIAM} for managing exponentially growing bases as exemplified for SIAM. We stress that the NN support in SOLAX is not a simple function containing the entire algorithm from Ref.~\cite{Bilous2024_SIAM}. Instead, we provide modular building blocks that users can easily customize and adapt to their specific research needs. We also demonstrate how these building blocks can be used to reconstruct the algorithm from Ref.~\cite{Bilous2024_SIAM}. To assist users with little or no prior experience in NNs, the chapter begins with an introduction into the basic NN concepts relevant to understand the code examples.

\subsection{Introduction to neural networks}

\subsubsection{Regression with dense neural networks\label{sec:nn_intro_regression}}

Although in the present work a NN is used for solution of the classification task, we start this introduction from considering the regression problem. This path is typically taken in literature since it allows to introduce many important concepts in a clear and intuitive way. On the other hand, as will be discussed below, NN-supported regression can be turned into NN-supported classification by a few modifications.

The regression task consists in approximating a multidimensional function \mbox{$f: \mathbb{R}^N \rightarrow \mathbb{R}^M$} given its values $\tilde{y}_1, \ldots, \tilde{y}_P$ on $P$ points $x_1, \ldots, x_P$. In this dataset, some noise may be present and each $\tilde{y}$ may slightly differ from the corresponding value $f(x)$. We start from NNs of the feedforward dense architecture which are the most general and conceptually simplest approximators for continuous functions. Such a NN consists of $L$ layers $i = 1, \ldots, L$ each representing the most general linear transformation $z^{(i)} = W^{(i)} x^{(i)} + b^{(i)}$ of the input vector $x^{(i)}$ with some matrix $W^{(i)}$ (kernel) and vector $b^{(i)}$ (bias). The output $z^{(i)}$ of each intermediate layer $i < L$ is additionally transformed with a non-linear function $h^{(i)}$ (e.g. the so called rectified linear unit, ReLU) and passed as input to the next layer: $x^{(i + 1)} = h^{(i)}\left(z^{(i)}\right)$. In the case of regression, we take $y^{(L)} = z^{(L)}$ for the last layer. With an appropriate selection of kernels and biases for the layers of the NN, the entire network serves as a transformation $y(x)$, mapping the input of the first layer $x = x^{(1)}$ to the output of the last layer $y = y^{(L)}$, thereby approximating the continuous functional relationship $f(x)$ based on the provided dataset.

\begin{figure}[!ht]
     \begin{center}
     \includegraphics[width=0.8\textwidth]{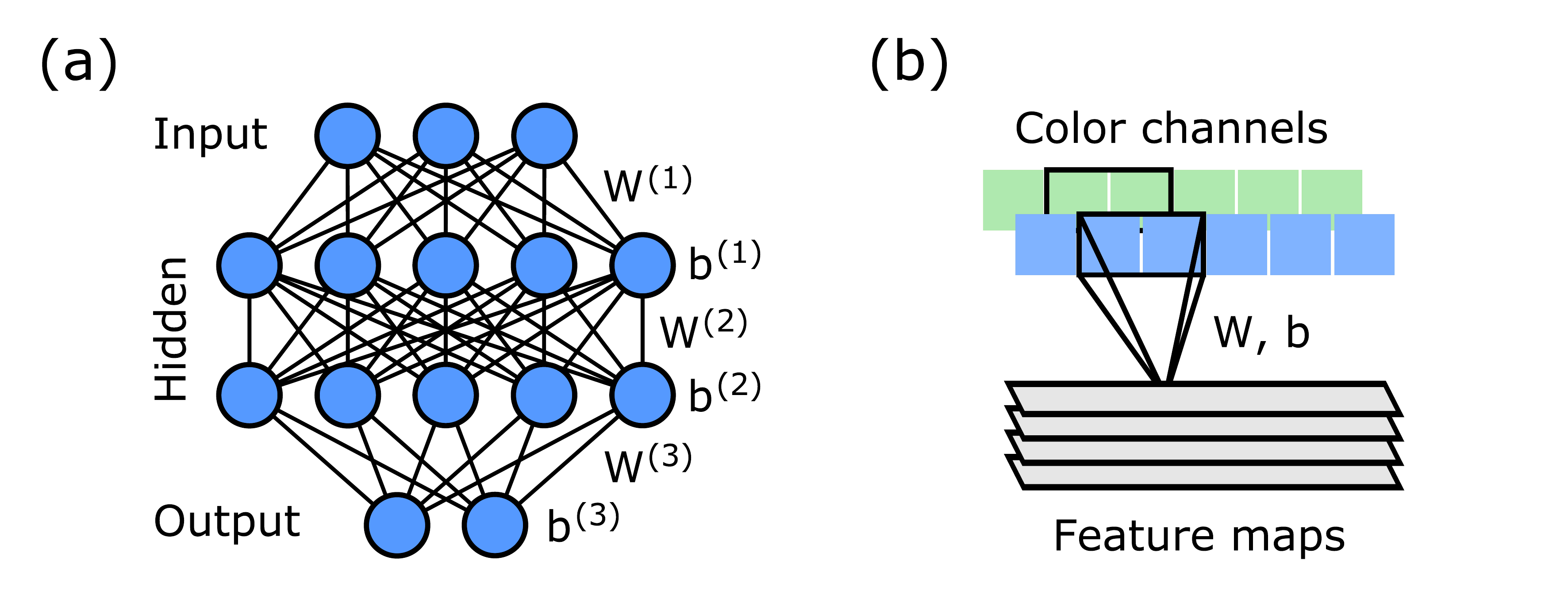}
     \end{center}
     \caption{(a) An example of a dense feedforward NN with $L=3$ kernel-bias pairs, and thus 2 hidden layers. (b) An example of a convolutional layer with $C=2$ color input channels, kernel length $D=2$, and $T=4$ output feature maps.}
     \label{fig:nn_intro}
\end{figure}

It is convenient to view the NN layers as consisting of nodes (neurons) each corresponding to one component of the output vector $z^{(i)}$. In Fig.~\ref{fig:nn_intro}(a) we show a schematic illustration of a dense feedforward NN with $L=3$. The kernel matrix $W^{(i)}$ is represented as connections between the layers $i$ and $i-1$, whereas the bias vector $b^{(i)}$ is associated with the neurons in each layer. In NNs of the usual dense architecture, each neuron is connected with all neurons of the previous layer. Note that the input $x = x^{(1)}$ to the whole NN is usually called ``input layer'' which, however, is not associated with any data transformation. The $L-1$ layers enclosed between the input and output layers are referred to as hidden layers.

\subsubsection{Neural network training}

Training of a NN is a process of iterative improvement of the unknown parameters $W^{(i)}$ and $b^{(i)}$ also referred to as trainable parameters. The NN performance is characterized by a loss function $l(y, \tilde{y})$ which for a given $x$ from the training dataset quantifies the mismatch between the NN-predicted value $y$ and the corresponding ``correct'' value $\tilde{y}$. Training is usually performed not on individual data entries $(x, \tilde{y})$, but on batches $(x_1, \tilde{y}_1), \ldots, (x_B, \tilde{y}_B)$ of a fixed size $B$. The loss on a batch is then the average over its entries:
\begin{equation}
    L = \frac{1}{B}\sum_{j = 1}^B l(y_j, \tilde{y_j})\;.
\end{equation}
In the case of regression, the typically used loss function is the square deviation
\begin{equation}
    l_2(y,\tilde{y}) = \frac{1}{2}(y - \tilde{y})^2\;.
\end{equation}

A training iteration on a data batch can be performed using the gradient descent approach as follows. First, the values $x_j$ with $j=1,\ldots,B$ are sent to the NN which predicts the values $y_j$. The loss $L$ on the batch is then computed as described above. In order to improve each NN trainable parameter which we denote here generically $\theta$, the gradients of the loss $\frac{\partial L}{\partial \theta}$ are evaluated. Each parameter $\theta$ is then improved towards the loss minimization as
\begin{equation}
    \theta_\mathrm{new} = \theta_\mathrm{old} - \eta \frac{\partial L}{\partial \theta}\;,
\end{equation}
where $\eta$ is the so called learning rate common for all trainable parameters. Note that $\eta$ is a hyperparameter, i.e. a parameter fixed by the user and not following from the training procedure. In practice, more advanced gradient-based update rules are often applied in order to stabilize and speed up convergence to the optimal values for the trainable parameters, e.g. the adaptive moment estimation (Adam) algorithm~\cite{kingma2017adam} used also in the present work. One key feature of this algorithm is that it uses adaptive learning rates for each parameter, based on the first moment (mean) and the second moment (variance) of the gradients.

Usually, NN training requires going batch by batch through the whole dataset multiple times (epochs). As convergence is achieved and no further improvement is observed, the training can be stopped. Such ``early stopping'' does not only speedup the training process, but also helps to avoid overfitting, which occurs when the NN learns the training data ``by heart'', leading to significantly worse performance on new, unseen data compared to the training data. To prevent overfitting, it is crucial to monitor the loss after each epoch not on the training set, but on a separate validation set held out from the dataset before training.

\subsubsection{Neural network as a classifier}

In the recent decades, the classification problem has been successfully tackled with NNs, in particular, in the domain of image recognition~\cite{NIPS2012_c399862d}. In the present work we demonstrate an algorithm which uses a NN classifier in an analogous way to iteratively select the most important Slater determinants. Here we show the modifications which turn the introduced NN regressor into a classifier used in this work.

In the classification problem, the training dataset must contain for each $x$ a discrete value $\tilde{y}$ indicating the correct class for $x$. We use here the one-hot encoding approach: If there are $K$ distinct classes, the $k$-th class is encoded by a vector $(0, \ldots, 1, \ldots, 0)$ of length $K$ where only the $k$-th component is non-zero and equal to 1. Each vector $\tilde{y}$ in one-hot encoding can be also interpreted as a set of probabilities for the corresponding $x$ to belong to each class. Since we have the full confidence about the correct class for each $x$ from the dataset, the probability distribution $\tilde{y}$ is degenerate. The corresponding approximative distribution $y(x)$ as predicted by the NN is, however, generally speaking non-degenerate. In this way, the inherently discrete-valued classification problem reduces to a problem with continuous values enabling us to use NNs similarly to the case of regression.

Whereas the hidden layers for a NN classifier can be constructed in the same way as in the case of a NN regressor, the output layer needs modification in order to ensure that the NN prediction $y(x)$ is a valid probability distribution. For the softmax-classifier considered here, the output vector $z=z^{(L)}$ of the last NN layer (see Section~\ref{sec:nn_intro_regression}) is sent through the softmax nonlinearity performed in two steps. Firstly, the NN output is made positive by element-wise exponentiation:
\begin{equation}
    u = \left(
    e^{z{[1]}}, \ldots, e^{z{[K]}}
    \right)\;,
\end{equation}
where $z[k]$ denotes the $k$-th component of the vector $z$. Secondly, the resulting vector is normalized ensuring additionally that the NN output sums to 1:
\begin{equation}
    y = \frac{u}{\sum_{k=1}^K u[k]}\;.
\end{equation}
Note that in contrast to e.g. ReLU which acts on the components of the vector argument independently (locally), the softmax operation is non-local due to the normalization step.

The natural measure quantifying the difference between the NN output probability distribution $y$ and the corresponding ``correct'' degenerate distribution $\tilde{y}$ is cross-entropy
\begin{equation}
    s(y, \tilde{y}) = -\sum_{k=1}^K \tilde{y}[k] \log y[k] \;,
\end{equation}
which is always non-negative and becomes zero for the targeted case $y = \tilde{y}$. Training for minimization of the cross-entropy loss is performed in a similar manner as in the case of regression with the square loss. After each training epoch, the classification accuracy (i.e. the fraction of the dataset entries classified by the NN correctly) can be additionally evaluated on a validation set and used for early stopping.

\subsubsection{Convolutional neural networks\label{sec:nn_intro_convol}}

Among other classification tasks, the image classification problem possesses a special property of translational invariance. For example the type of a vehicle shown in an image does not depend on where exactly it is placed with respect to the image frame. This symmetry can be accounted for directly in the NN architecture by using convolutional layers prior to a dense feedforward classifier. We formulate here at the general level the concepts from convolutional NNs relevant to our work.

Whereas image recognition deals usually with 2D images, we concentrate here on 1D ``images'' (which in the present work encode Slater determinants). A convolutional layer [see Fig.~\ref{fig:nn_intro}(b)] applies a linear transformation to small ``windows'' of the input pixels $x[d], \ldots, x[d + D - 1]$ of a fixed length $D$ and different starting positions $d$. For each $d$, the transformation returns a scalar value
\begin{equation}
    z[d] = \sum_{j=1}^D W[j] \cdot x[j + d - 1] + b\;,
\end{equation}
where the kernel vector $W$ of length $D$ and the bias scalar $b$ are trainable parameters. As $d$ is varied and the window is moved within the image, the transformation results in a converted image typically referred to as a feature map. Note that for $D > 1$ the obtained feature map is smaller than the original image. In the image recognition domain, the same size for the feature map and the original image is often enforced by padding the latter with corresponding number of zeros beyond its frame. In the present work, only the actual data are used and no padding is performed.

If a few color channels indexed by $c = 1, \ldots, C$ are present, a feature map is constructed as
\begin{equation}
    z[d] = \sum_{c=1}^C\sum_{j=1}^D W[c,j] \cdot x[c,j + d - 1] + b\;.
\end{equation}
Usually dozens of feature maps are constructed at the same time as
\begin{equation}
    z[t,d] = \sum_{c=1}^C\sum_{j=1}^D W[t,c,j] \cdot x[c,j + d - 1] + b[t]\;,
\end{equation}
where the new index $t = 1,\ldots, T$ labels the feature maps. The layer output $y[t,d]$, which is obtained from $z[t,d]$ by additionally applying a non-linear transformation (e.g. ReLU), can serve as input with $T$ ``color'' channels to the next convolutional layer, or can be flattened an forwarded to a dense NN classifier.

We stress that convolutional NNs actually used for image recognition usually contain further transformations (e.g. pooling and upsampling) and additional variations of the architecture, see e.g. Ref.~\cite{Goodfellow2016}. We only provided here the necessary information for demonstration of the algorithm from Ref.~\cite{Bilous2024_SIAM} for managing large sets of Slater determinants.

\subsection{Algorithm description\label{sec:algo_core}}

\begin{figure}[!ht]
     \begin{center}
     \includegraphics[width=0.7\textwidth]{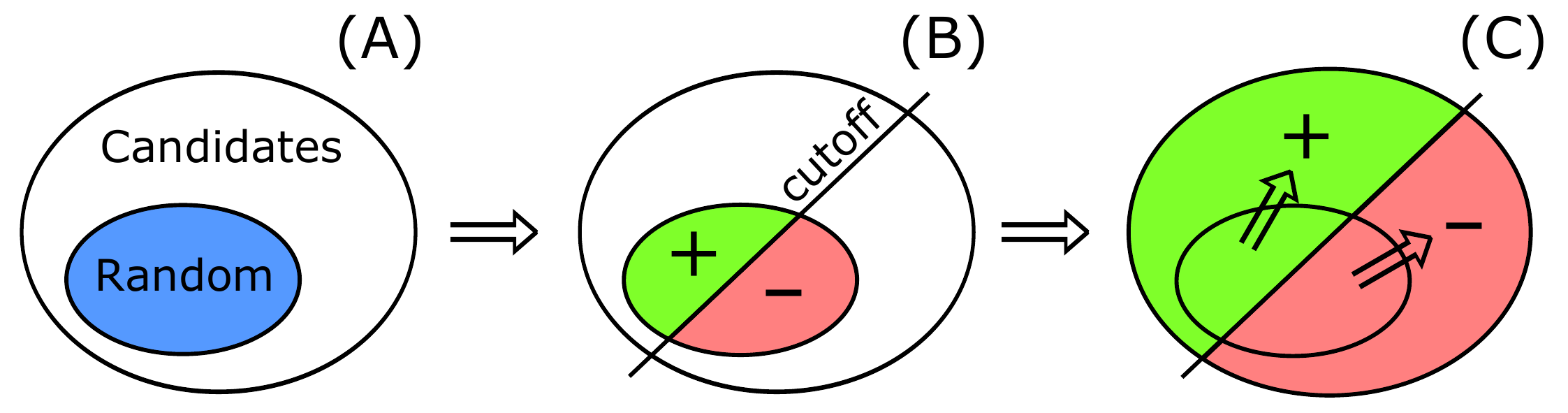}
     \end{center}
     \caption{Schematic illustration of the method developed in Ref.~\cite{Bilous2024_SIAM}. Only the core part of the algorithm is shown. See text for details.}
     \label{fig:algo_core}
\end{figure}

Following this introductory section on machine learning, we will now outline the core concept of the algorithm introduced in Ref.~\cite{Bilous2024_SIAM}, designed to handle large basis sets employing a NN classifier. A schematic illustration of the algorithm is provided in Fig.~\ref{fig:algo_core}. When a basis extension generates an excessive number of new Slater determinants, not all of these new ``candidates'' are included in the calculation, but only a fraction of \textit{the most important} ones. The importance of a determinant is quantified by its weight, which corresponds to the magnitude of its expansion coefficient in the quantum state. The fraction $\alpha$ of determinants to be included is a parameter chosen by the user.

To fulfill the user's request, the following steps are performed. First, as illustrated in panel (A) of Fig.~\ref{fig:algo_core}, a random selection is drawn from the set of candidate determinants and added to the existing basis used in the computation (the existing basis is not shown in Fig.~\ref{fig:algo_core}). A diagonalization is then performed, providing the expansion coefficients for the randomly added determinants. Note that the size of the random selection is determined by the user.

Next, as illustrated in panel (B), a cutoff is chosen to divide the random selection into two classes: determinants with weights above the cutoff (``important'') and those below it (``unimportant''). Based on the known weights of the randomly selected determinants, the cutoff is automatically adjusted such that the important class comprises a fraction $\alpha$ of the random selection. If the algorithm is well applicable to the case at hand, this cutoff divides also the full set of candidates in a similar proportion (note that the algorithm performance should be investigated for each specific case e.g. as we demonstrate in Section~\ref{sec:nn_check_results}). It remains unknown \textit{which} determinants outside the random selection belong to each class, and this is where a NN is useful.

At this stage, as shown in panel (C), a NN classifier is employed to categorize the remaining determinants into the importance classes. The random selection serves as the training data. For each determinant in the training set, the spin-orbital occupations are used as input features for the NN, while the determinant class serves as the "correct answer". Once trained, the NN is applied to the rest of the candidates to predict their importance class. This process sorts out the entire set of candidates, allowing only the important ones to be retained for diagonalization. We focus here on demonstrating this core procedure. Using the presented SOLAX tools, the algorithm can be further modified (e.g., in Ref.~\cite{Bilous2024_SIAM}, some determinants outside the set of candidates were also used for NN training).

In SOLAX, we implemented a custom machine learning package based on the FLAX library~\cite{flax2020github} from the JAX ecosystem. This package is primarily intended for development purposes and is not exposed to the user at the general SOLAX interface level. We employed these machine learning tools to provide the necessary functionality for implementing algorithms such as the one described. This functionality is encapsulated in two classes available at the SOLAX interface level: \verb+BasisClassifier+ and \verb+BigBasisManager+. The former encapsulates a NN setting, while the latter contains methods that reflect parts of the described algorithm. For demonstration, we continue with the SIAM example computation of Section~\ref{sec:siam_noml}.

\subsection{BasisClassifier\label{sec:basis_classifier}}

We now turn to a discussion of the \verb+BasisClassifier+ class, which integrates a NN and an optimizer to facilitate necessary machine learning operations. 
The user is expected to create a NN architecture using FLAX building blocks, while an optimizer is selected from the Optax package available alongside JAX. First, we make the necessary imports:
\begin{minted}{python}
from flax import linen as nn
import optax
\end{minted}

\noindent Here, the \verb+flax.linen+ API is imported via the conventional name \verb+nn+. A NN architecture can be now defined using the standard FLAX approach, i.e. via writing a Python function describing how a single data entry propagates towards the classification output. In this function, the FLAX building blocks as well as general JAX transformations can be used. We implement here the convolutional architecture employed in the work~\cite{Bilous2024_SIAM}, which was at the time implemented based on the TensorFlow library~\cite{TensorFlow2015}.
\begin{minted}{python}
def nn_call_on_bits(x):
    x = x.reshape(-1, 2)
    x = nn.Conv(features=64, kernel_size=(2,), padding="valid")(x)
    x = nn.relu(x)
    x = nn.Conv(features=4, kernel_size=(1,), padding="valid")(x)
    x = nn.relu(x)
    x = x.reshape(-1)

    x = nn.Dense(features=dense_size)(x)
    x = nn.relu(x)
    x = nn.Dense(features=dense_size//2)(x)
    x = nn.relu(x)
    x = nn.Dense(features=dense_size//4)(x)
    x = nn.relu(x)
    x = nn.Dense(features=2)(x)
    return x
\end{minted}
\begin{figure}[ht!]
     \begin{center}     \includegraphics[width=0.7\columnwidth]{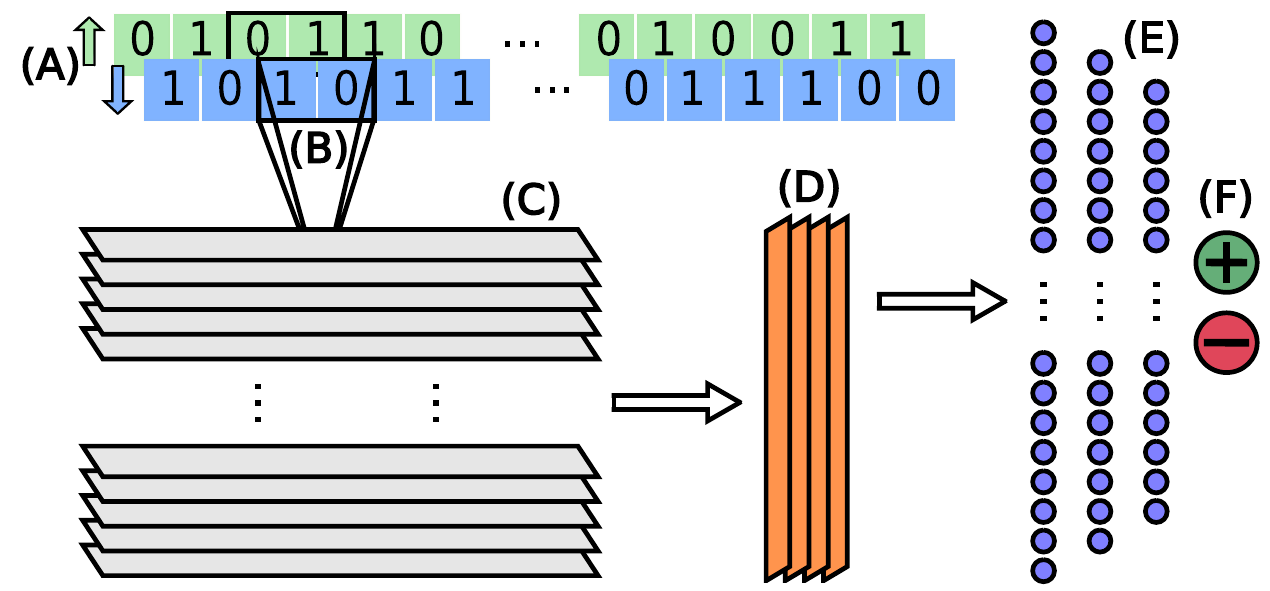}
     \end{center}
     \caption{Architecture of the convolutional NN as implemented in the function \texttt{nn$\_$call$\_$on$\_$bits}. 
     A candidate Slater determinant is encoded as spin-orbital populations distributed into two spin-channels (A). This input is processed by a convolutional kernel (B) of size 2 into 64 feature maps (C), and then by a convolutional kernel of size 1 (now shown in the plot) into 4 feature maps (D). Both convolutional layers have ReLU activation. The output is flattened and sent to the dense block (E)---(F) which ends with two classification logits (F) classifying the determinant as important/unimportant. Note that \texttt{nn$\_$call$\_$on$\_$bits} does not contain the softmax activation, since SOLAX automatically applies softmax to the output logits (F).
     %
     }
     \label{fig:nn_architecture}
\end{figure}
\noindent We depict this function schematically in Fig.~\ref{fig:nn_architecture} labeling the relevant parts using capital letters (A)---(F). The input \verb+x+ is an array of bits (0 and 1) representing the spin-orbital populations of the incoming Slater determinant. The input data are distributed into the spin-up ($\uparrow$) and spin-down ($\downarrow$) channels (A) using \verb+reshape+. This representation is processed by a convolutional kernel (B) of size 2 into 64 feature maps (C), and then by a convolutional kernel of size 1 (now shown in the plot) into 4 feature maps (D). Both convolutional layers have ReLU activation. The argument \verb+padding="valid"+ indicates that there is no additional padding with zeros, see Section~\ref{sec:nn_intro_convol}. The output of the convolutional block (A)---(D) is flattened using \verb+reshape(-1)+ and passed to the dense block (E)---(F), which ends with two neurons (F) containing the classification logits. Note that in the code calling this function, the output logits are automatically passed through the softmax activation function to convert them into probabilities summing to 1. Hence, there is no need to explicitly apply the softmax function within the NN definition.

In the function body above, the size of the layers in the dense block is determined by the global variable \verb+dense_size+
initialized as follows:
\begin{minted}{python}
dense_size = int(7 * np.sqrt(2 * N_bath + 2))
print(dense_size)
\end{minted}
\begin{verbatim}
    46
\end{verbatim}
Here, we provided the generic expression for arbitrary $N_\text{bath}$ used in Ref.~\cite{Bilous2024_SIAM}. Empirically, this turned out to be a good choice in the case of SIAM. Also for other models it could serve as a reasonable starting point for further search and fine-tuning. An object of \verb+BasisClassifier+ is now instantiated as
\begin{minted}{python}
classifier = sx.BasisClassifier(nn_call_on_bits)
\end{minted}
\noindent We note that so far no NN has been actually initialized in memory. The latter needs to be performed explicitly using the \verb+initialize+ method which requires the following additional arguments: (1) an input \verb+Basis+ prototype, (2) an optimizer, and (3) a JAX random key.

The input \verb+Basis+ prototype required by the \verb+BasisClassifier+ \verb+initialize+ method is needed to find out the size of the input \verb+x+ of the function defining the NN. This determines also the concrete structure of the NN as created in memory. We use here the variable \verb+basis_start+ from the SIAM example in Section~\ref{sec:siam_noml}. We choose the standard Adam optimizer~\cite{kingma2017adam} which can be created using Optax as
\begin{minted}{python}
optimizer = optax.adam(learning_rate=0.005)
\end{minted}

\subsubsection{RandomKeys}

In JAX, randomization is performed in a deterministic and reproducible manner. Each function generating random numbers receives a JAX random key which determines the randomization. These keys can be created from each other, whereas the very first one is created from an integer seed. We wrapped this mechanism in a convenience class \verb+RandomKeys+ which implements the Python iterator interface. A \verb+RandomKeys+ instance is created as
\begin{minted}{python}
rand_keys = sx.RandomKeys(seed=1234)
\end{minted}
\noindent where \verb+seed+ is a keyword-only argument. Now, each time when a new JAX random key needs to be generated, the Python \verb+next+ keyword can be used on \verb+rand_keys+:
\begin{minted}{python}
key_for_nn = next(rand_keys)
\end{minted}
\noindent The obtained key is passed to the \verb+initialize+ method in order to randomly initialize the NN weights.

\vspace{0.75\baselineskip}
\noindent Using the needed components, we initialize the NN in memory as
\begin{minted}{python}
classifier.initialize(key_for_nn, basis_start, optimizer)
\end{minted}
\noindent Note that once initialized, the NN summary can be printed out using the \verb+print_summary+ method (we skip this step here due to the output size). This exhausts the knowledge about the \verb+BasisClassifier+ class needed to proceed with implementation of the sketched algorithm using \verb+BigBasisManager+.

\subsection{BigBasisManager}

The central SOLAX tool for implementing algorithms as the one described above, is the \\\verb+BigBasisManager+ class. Its objects are created from:
\begin{itemize}
\item a set of candidate Slater determinants represented as an object of the SOLAX class \verb+Basis+;
\item a NN setting represented as an object of the just shown SOLAX class \verb+BasisClassifier+.
\end{itemize}
In order to demonstrate the functions provided with \verb+BigBasisManager+, we turn back to the SIAM example with $N_\text{bath} = 21$ bath sites considered in Section~\ref{sec:siam_noml_full_comp}. In the iterative solution shown there, we achieved the basis size of 7084 Slater determinants. Another extension yields
\begin{minted}{python}
basis_small = basis

basis_big = H(basis_small)
print(len(basis_big))
\end{minted}
\begin{verbatim}
    58984
\end{verbatim}
Here we stick to the notations adopted also in Section~\ref{sec:siam_noml_matrix_optim}. The obtained basis poses no computational challenge and was tackled directly in Section~\ref{sec:siam_noml_matrix_optim}. Still, for demonstration, we sort it out using the  NN-supported algorithm and compare the resulting state energy with the one obtained directly. The set of new ``candidates'' generated at the last extension step is obtained as
\begin{minted}{python}
candidates = basis_big % basis_small
print(len(candidates))
\end{minted}
\begin{verbatim}
    51900
\end{verbatim}
Using also the \verb+classifier+ object built above, we create now a \verb+BigBasisManager+ instance:
\begin{minted}{python}
bbm = sx.BigBasisManager(candidates, classifier)
\end{minted}
\noindent Instead of including all 51900 candidates in the computation, we follow Ref.~\cite{Bilous2024_SIAM} and target inclusion of the following number of \textit{the most important} determinants:
\begin{minted}{python}
target_num = int(np.sqrt(len(basis_big)) * 50)
print(target_num)
\end{minted}
\begin{verbatim}
    12143
\end{verbatim}
We provided here the empirical generic expression used in Ref.~\cite{Bilous2024_SIAM} for SIAM, which for other systems can be used as a reasonable starting point for further search. In the following we implement the sketched NN-based approach with the help of the created \verb+bbm+ object.

\subsubsection{Random selection}

For the random selection size, we follow again the empirical choice from Ref.~\cite{Bilous2024_SIAM} for SIAM, and use the following expression for the size of the random selection shown in Fig.~\ref{fig:algo_core}(A):
\begin{minted}{python}
random_num = int(target_num / 1.5)
print(random_num)
\end{minted}
\begin{verbatim}
    8095
\end{verbatim}
The selection can be drawn from \verb+candidates+ using the \verb+bbm+ object via the \verb+sample_subbasis+ method:
\begin{minted}{python}
random_sel = bbm.sample_subbasis(next(rand_keys), random_num)
\end{minted}
\noindent Here we generated another JAX random key from the \verb+rand_keys+ object and passed it directly to the \verb+sample_subbasis+ function. The returned \verb+random_sel+ object is of type \verb+Basis+. In order to obtain the weights for the picked determinants as required by the algorithm, we add the random selection on top of the old basis:
\begin{minted}{python}
basis_diag = basis_small + random_sel
print(len(basis_diag))
\end{minted}
\begin{verbatim}
    15179
\end{verbatim}
and perform diagonalization as shown in Section~\ref{sec:siam_noml_full_comp}:
\begin{minted}{python}
matrix = H.build_matrix(basis_diag)
result = sp.sparse.linalg.eigsh(matrix.to_scipy(), k=1, which="SA")

energy = result[0][0]
print(f"Intermediate energy:\t{energy}")
\end{minted}
\begin{verbatim}
    Intermediate energy:	-31.720920015599123
\end{verbatim}
Finally, we use the obtained eigenvector to create a \verb+State+:
\begin{minted}{python}
eigenvec = result[1][:, 0]
state_diag = sx.State(basis_diag, eigenvec)
\end{minted}
\noindent from which we strip \verb+basis_small+ obtaining the randomly selected determinants together with their coefficients encapsulated in a \verb+State+ object:
\begin{minted}{python}
state_train = state_diag % basis_small
\end{minted}
\noindent Note that the latter object serves only for convenient storing the random selection coefficient and does not actually represent the searched quantum state. The variable name \verb+state_train+ reflects that this \verb+State+ instance contains the data which will be used for the NN training.

As expected, the intermediate energy obtained above lies between the energy on \verb+basis_small+ computed in Section~\ref{sec:siam_noml_full_comp} and the energy on \verb+basis_big+ from Section~\ref{sec:siam_noml_matrix_optim}. The corresponding separations are approx. $0.0136$ and $0.0981$. In this way, inclusion of the random selection in the computation does not considerably promote the energy towards the value on the larger basis. In contrast, as we will see in the following sections, inclusion of NN-selected determinants pushes the energy close to the value on \verb+basis_big+.

\noindent

\subsubsection{Deriving the cutoff}

We switch now to splitting the random selection with a cutoff as discussed in Section~\ref{sec:algo_core} and illustrated in Fig.~\ref{fig:algo_core}(B). The cutoff can be obtained using the \verb+BigBasisManager+ method \verb+derive_abs_coeff_cut+:
\begin{minted}{python}
abs_coeff_cut = bbm.derive_abs_coeff_cut(target_num, state_train)
print(f"Cutoff:\t{abs_coeff_cut}")
\end{minted}
\begin{verbatim}
    Cutoff:	0.00020762079204639022
\end{verbatim}
This cutoff splits the random selection into two importance classes of determinants with larger and smaller weights. We point out again that under weight we understand here the absolute value of the determinant coefficient (not its square). The ``important'' class as represented by a \verb+State+ object is obtained by chopping
\begin{minted}{python}
state_train_impt = state_train.chop(abs_coeff_cut)
print(len(state_train_impt))
\end{minted}
\begin{verbatim}
    1893
\end{verbatim}
The cutoff is automatically chosen such that the fraction of the important class in the random selection
\begin{minted}{python}
print(len(state_train_impt) / len(state_train))
\end{minted}
\begin{verbatim}
    0.23384805435453984
\end{verbatim}
equals the ratio of the targeted number of the most important determinants to be included to the full number of the candidates
\begin{minted}{python}
print(target_num / len(candidates))
\end{minted}
\begin{verbatim}
    0.23396917148362234
\end{verbatim}
%

\subsubsection{Using the neural network}

In the following, we employ the NN in order to distribute the determinants outside the random selection into the importance classes as illustrated in Fig.~\ref{fig:algo_core}(C). The NN is already created, initialized and incorporated in the \verb+bbm+ object of the \verb+BigBasisManager+ class. The latter provides high-level functions for the NN usage in the considered context. We train the NN similarly to Ref.~\cite{Bilous2024_SIAM} as follows:
\begin{minted}{python}
early_stopped = bbm.train_classifier(
    next(rand_keys),
    state_train,
    abs_coeff_cut,
    batch_size=256,
    epochs=200,
    early_stop=True,
    early_stop_params={"patience": 3}
)
\end{minted}
\begin{verbatim}
    Started:	accuracy=2.472703e-01
    Epoch 0:	accuracy=8.015612e-01
    Epoch 1:	accuracy=8.385798e-01
    Epoch 2:	accuracy=8.525778e-01
    Epoch 3:	accuracy=8.595095e-01
    Epoch 4:	accuracy=8.705761e-01
    Epoch 5:	accuracy=8.619702e-01
    Epoch 6:	accuracy=9.055037e-01
    Epoch 7:	accuracy=8.985720e-01
    Epoch 8:	accuracy=9.237778e-01
    Epoch 9:	accuracy=9.326122e-01
    Epoch 10:	accuracy=9.412651e-01
    Epoch 11:	accuracy=9.312204e-01
    Epoch 12:	accuracy=9.491246e-01
    Epoch 13:	accuracy=9.461933e-01
    Epoch 14:	accuracy=9.596332e-01
    Epoch 15:	accuracy=9.524729e-01
    Epoch 16:	accuracy=9.450301e-01
    Epoch 17:	accuracy=9.506576e-01
    Epoch 18:	accuracy=9.534478e-01
\end{verbatim}
Here, a new JAX random key is generated from \verb+rand_keys+ and passed directly to the \\\verb+train_classifier+ call. It is needed for reshuffling to avoid any ordering bias in the training data. The latter are provided as the \verb+state_train+ object of the \verb+State+ class. Whereas \verb+state_train+ contains only the Slater determinants with their coefficients, the \\\verb+train_classifier+ method takes care of converting this to classification data based on the cutoff \verb+abs_coeff_cut+ also provided to the method.

The rest of the parameters are keyword-only arguments controlling the NN training, which is performed in batches of size 256 until the best performance is achieved and then early-stopped. After the epoch in which the NN failed to improve, we give it a chance to try 3 more times by indicating \verb+patience=3+. The early stopping parameters passed as the \\\verb+early_stop_params+ dictionary are forwarded directly to the underlying \verb+EarlyStopping+ FLAX class and can be looked up in the FLAX documentation~\cite{flax2020github}. Note that the NN is reset to its best state achieved in the training process. The output value \verb+early_stopped+ is \verb+True+ if the training was early-stopped, and \verb+False+ if it went through all 200 epochs (provided with the \verb+epochs+ argument) without reaching the best performance.

The printed NN training information contains the NN classification accuracy evaluated before the training and after each epoch on a data part held out from the training set. The fraction of the data used for the performance evaluation can be controlled via the keyword-only argument \verb+val_frac+ of the \verb+train_classifier+ method (the default value is \verb+val_frac=0.2+). We note that if an NVIDIA GPU is available and a GPU-capable version of JAX is installed on the machine, it will be automatically used for the NN training usually leading to a significant speedup.

The trained NN can be now used for classification of all candidates:
\begin{minted}{python}
nn_selected = bbm.predict_impt_subbasis(batch_size=256)
nn_selected = nn_selected % state_train.basis
print(len(nn_selected))
\end{minted}
\begin{verbatim}
    9834
\end{verbatim}
The method \verb+predict_impt_subbasis+ returns a \verb+Basis+ object \verb+nn_selected+ containing the Slater determinants classified by the NN as important. Note that the NN is applied here to the full set of candidates including the training subset present in the \verb+State+ object \verb+state_train+. Therefore, we strip the latter from \verb+nn_selected+. The NN prediction operation runs automatically on a GPU if available. In this case, the keyword-only \verb+batch_size+ argument can be passed to the \verb+predict_impt_subbasis+ method for controlling the GPU memory usage.

The final subset of candidates to be included in the computation as the result of the NN-supported procedure is build as
\begin{minted}{python}
basis_impt = nn_selected + state_train_impt.basis
print(len(basis_impt))
print(abs(len(basis_impt) - target_num) / target_num)
\end{minted}
\begin{verbatim}
    11727
    0.034258420489170716
\end{verbatim}
Here we printed the size of the obtained important subset and its relative deviation from the targeted size \verb+target_num+. We see that the user's request has been satisfied in this demonstration example.

\subsubsection{Checking and processing of the results\label{sec:nn_check_results}}

We construct now the full basis and evaluate the state energy:
\begin{minted}{python}
basis = basis_small + basis_impt

matrix = H.build_matrix(basis)
result = sp.sparse.linalg.eigsh(matrix.to_scipy(), k=1, which="SA")

energy = result[0][0]
print(f"Basis:\t{len(basis)}")
print(f"Energy:\t{energy}")
\end{minted}
\begin{verbatim}
    Basis:	18811
    Energy:	-31.817043901573747
\end{verbatim}
The obtained energy is separated from the energy computed on \verb+basis_small+ in Section~\ref{sec:siam_noml_full_comp} and the energy on \verb+basis_big+ from Section~\ref{sec:siam_noml_matrix_optim} by $0.1100$ and $0.0017$, respectively. In this way, by using the NN support provided in SOLAX we were able to almost reach the same state energy on 19462 Slater determinants instead of 58984.

We use now the obtained eigenvector to check how many determinants out of those suggested by the NN indeed possess weights higher that \verb+abs_coeff_cut+. First we construct a \verb+State+ object corresponding to the NN suggestion:
\begin{minted}{python}
eigenvec = result[1][:, 0]
state = sx.State(basis, eigenvec)

nn_selected_state = state % basis_small % state_train.basis
print(nn_selected_state.basis == nn_selected)
\end{minted}
\begin{verbatim}
    True
\end{verbatim}
The \verb+State+ instance \verb+nn_selected_state+ contains a basis set equal to \verb+nn_selected+ and additionally the evaluated coefficients. Now we chop off the misclassified determinants:
\begin{minted}{python}
nn_selected_right = nn_selected_state.chop(abs_coeff_cut).basis
print(len(nn_selected_right))
print(len(nn_selected_right) / len(nn_selected))
\end{minted}
\begin{verbatim}
    8648
    0.8793980069147854
\end{verbatim}
The printed fraction of the Slater determinants classified correctly is smaller that the NN accuracy achieved in the training procedure. We attribute this to the drift of the determinant coefficients upon the inclusion in the computation of other determinants. We stress again that the NN performance should be investigated individually for each application case e.g. following the outlined procedure.

If the obtained basis is involved in further computations (e.g. as the starting point of the next NN-supported iteration), it is advantageous to exclude from it the misclassified determinants:
\begin{minted}{python}
nn_selected_wrong = nn_selected % nn_selected_right
basis_final = basis % nn_selected_wrong
print(len(basis_final))
\end{minted}
\begin{verbatim}
    17625
\end{verbatim}
If needed, the energy and the coefficients can be evaluated on \verb+basis_final+ via the corresponding Hamiltonian matrix. Note, however, that it is inefficient to compute the latter from scratch using the \verb+H.build_matrix+ call. Instead, as demonstrated in the section on the \verb+OperatorMatrix+ class, the \verb+shrink_basis+ method of the \verb+matrix+ object can be used to extract the matrix on \verb+basis_final+ directly from the \verb+OperatorMatrix+ instance \verb+matrix+ built on \verb+basis+. This shortcut is possible since \verb+basis_final+ represents a subset of \verb+basis+, and therefore all the needed matrix elements have been already evaluated.

We point out that each \verb+BigBasisManager+ instance is bound to a particular \verb+Basis+ object with candidates to be sorted out. Therefore, it is necessary to create a new \verb+BigBasisManager+ object for each new big basis to be optimized (e.g. if further NN-supported iterations follow). At the same time, the same \verb+BasisClassifier+ object can be reused as a component in different \verb+BigBasisManager+ instances transferring in this way the NN experience from case to case.

\noindent\\
To summarize, using the NN support tools provided in SOLAX, we could reach the same accuracy level for the SIAM ground state energy on a much smaller basis. Further iterations of the described algorithm would profit strongly from the reduced basis size as the starting point. In Ref.~\cite{Bilous2024_SIAM}, this algorithm was implemented using the TensorFlow library~\cite{TensorFlow2015} and the Quanty code~\cite{Lu2014} for working with fermionic quantum systems. It was applied to SIAM with up to $N_\text{bath} = 299$ bath sites to sort out many millions of Slater determinants and helped to reduce the necessary basis set sizes by orders of magnitude. As implemented in SOLAX, this approach was applied for the first time in Ref.~\cite{Schmerwitz2024_N2} for computations of the ground state of the N$_2$ molecule. Note that whereas we prefer NNs due to their scalability, flexibility and availability of powerful frameworks like JAX and TensorFlow, also other classifier types could be potentially employed in the described algorithm, as demonstrated in Refs~\cite{Jeong_ALCI_JChemTC_2021, Chembot} for similar computations. The limitation of the presented method can be considered in a two-fold way. From the conceptual perspective, its applicability to a particular problem is itself a research question, which can be addressed with the SOLAX tools we provide. From the technical perspective, i.e. in the context of the needed computational resources, it is important to address the computational time needed by different parts of the algorithm. In the following section we provide such benchmarks for larger NN-supported SIAM computations with GPU acceleration and purely on CPU.

\subsubsection{Computation time benchmarks}

\begin{figure}[h!]
    \centering
    \includegraphics[width=\linewidth]{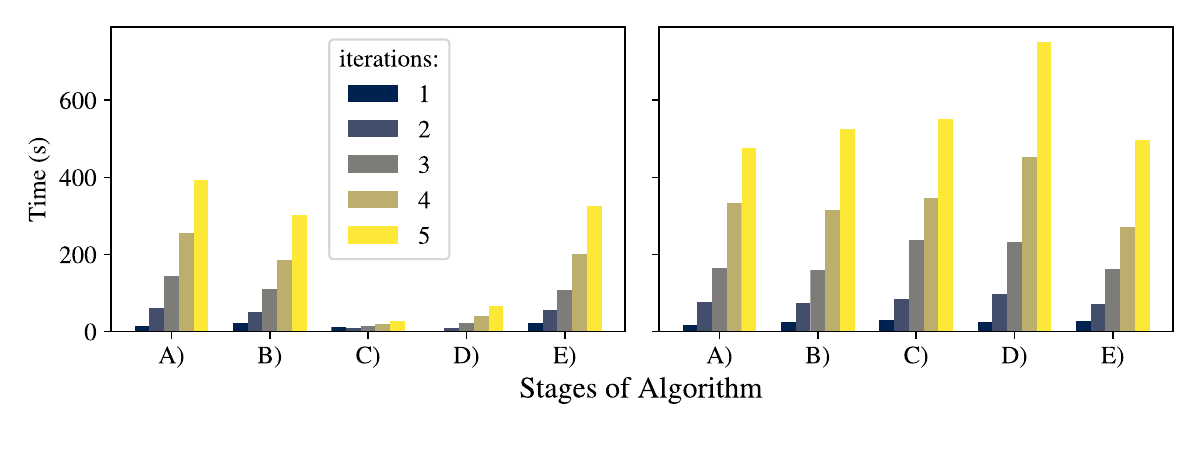}
    \caption{Comparison of time spent on different parts of our algorithm for SIAM with $N_\mathrm{bath} = 149$ as performed with single-GPU acceleration (left panel) and purely on CPU (right panel). The categories A)---E) are explained in the text.}
    \label{fig:time_analysis}
\end{figure}

In this section, we show comparisons of computation times spent in different parts of the described algorithm.  These comparisons cannot be performed in a representative way for such small examples as presented above. Therefore, we turn here to more extensive computations for the SIAM with 149 bath sites and perform five NN-supported iterations. One iteration of the NN-supported algorithm can be divided into five main components listed below. Note that we follow here exactly Ref.~\cite{Bilous2024_SIAM} where a slightly modified algorithm was employed. Whereas the first NN-supported iteration is identical with the algorithm version described above, in further iterations modifications are included which are specified below for the individual algorithm components.
\begin{enumerate}[label=\Alph*)]
    \item \textbf{Basis Extension:} This step involves applying an extension operator to \verb+basis_small+ to generate new candidate Slater determinants. Starting from the second iteration, we form the new set of candidates not by directly acting with the extension operator on the set of determinants currently present in the computation, but by acting only on the determinants included in the last iteration, and combining the result with the set of candidates from the last iteration. Whereas the obtained set is the same, the latter procedure is less costly, see Ref.~\cite{Bilous2024_SIAM}.
    \item \textbf{Training Data Generation:} This process selects \verb+random_num+ determinants randomly, performs a partial diagonalization of the Hamiltonian represented on \\\verb|basis_small + random_sel|, and computes the cutoff coefficient to label the determinants in \verb+random_sel+. Beyond the first iteration, the Slater determinants included in the previous NN-supported iterations are additionally used for the NN training, see Ref.~\cite{Bilous2024_SIAM}.
    \item \textbf{Neural Network Training:} The NN classifier is trained using the generated training data.
    \item \textbf{Neural Network Prediction:} The trained NN predicts which of the candidate determinants are important,  resulting in the basis set \verb+nn_selected+.
    \item \textbf{Selection Correction:} To complete the iteration, a second partial diagonalization is performed on \verb|basis_small + nn_selected| along with the training determinants labeled as important.
\end{enumerate}
In Fig.~\ref{fig:time_analysis}, we present the time spent in these five algorithm parts with single-GPU acceleration and purely on CPU.  It is seen that the strongest GPU speedup is achieved in parts C) and D), i.e.  the NN training and prediction. The full computation on a cluster with a GPU took 0.7 hours, and on a CPU cluster 1.7 hours. The same computation performed in Ref.~\cite{Bilous2024_SIAM} on a CPU cluster took about 35 hours dominated by the inefficient communication between the Quanty CI code and the Python code with a NN.

We note that for more complex Hamiltonians,  e.g. for atoms and molecules including inter-orbital interactions, the ratio between the execution times for the algorithm parts can further shift towards the non-NN parts. For example, in computations performed for the N$_2$ molecule in Ref.~\cite{Schmerwitz2024_N2}, we observed that steps B) and E) involving the construction of an \verb+OperatorMatrix+ were significantly more expensive than the rest of the algorithm. At the same time, however, the observed GPU acceleration was here more pronounced than for the considered SIAM example.

\section{Saving/loading SOLAX objects and reproducing computations\label{sec:save_load}}

In this section, we focus on the crucial functionality of saving and loading objects of the SOLAX classes. This feature enables checkpointing computations and restoring them in a flexible and efficient way. SOLAX provides a convenient and unified mechanism for this purpose through the global functions \verb+save+ and \verb+load+, which are compatible with objects of the \verb+Basis+, \verb+State+, \verb+OperatorTerm+, \verb+Operator+, \verb+OperatorMatrix+, and \verb+RandomKeys+ classes. It is important to note that \verb+RandomKeys+ objects are saved in their current iteration state, rather than JAX random keys generated by them. The \verb+BasisClassifier+ class employs its own approach, which is based on the Orbax library~\cite{orbax} from the JAX ecosystem.

\subsection{Standard mechanism}

We start from the standard mechanism based on the \verb+save+ and \verb+load+ functions. Using this approach, it is possible to save and load objects of the SOLAX classes listed above. For instance, we can save the \verb+basis_final+ object containing the basis set of Slater determinants built using the NN-supported algorithm in the previous sections:
\begin{minted}{python}
sx.save(basis_final, "solax_basis_")
\end{minted}
\noindent and load it again as
\begin{minted}{python}
basis_loaded = sx.load("solax_basis_")
print(basis_loaded == basis_final)
\end{minted}
\begin{verbatim}
    True
\end{verbatim}
Here we ensured that the loaded \verb+Basis+ object is indeed equal to the saved one. The string \verb+"solax_basis_"+ indicates the name of the directory we save to and load from.

\paragraph{Important note!} The directory is created prior to saving. If the directory already exists, it will be first erased together with its current content. Note that this applies also to the \verb+BasisClassifier+ class which has its own saving/loading mechanism (see below).

Instead of saving and loading standalone SOLAX objects, it is possible to first bundle many objects in one Python dictionary. The user provides here a string label to each object as the key in a key-value pair, whereas the object itself is the value. For instance, for the objects \verb+basis_final+ and \verb+H+ from the previous section we use the string labels \verb+"basis_from_nn"+ and \verb+"hamiltonian"+, respectively:
\begin{minted}{python}
dict_to_save = dict(
    basis_from_nn=basis_final,
    hamiltonian=H
)

sx.save(dict_to_save, "solax_basis_ham_")
\end{minted}
\begin{minted}{python}
loaded_dict = sx.load("solax_basis_ham_")

for key, value in loaded_dict.items():
    print(f"{key} has type {type(value).__name__}")
\end{minted}
\begin{verbatim}
    basis_from_nn has type Basis
    hamiltonian has type Operator
\end{verbatim}
We printed here only the types of the loaded objects. The used key strings are employed in internal addressing of the saved objects and must satisfy the following rule: A key string is valid if it \textit{could be} used as a Python variable name. Therefore,  we recommend to create dictionaries for saving using the \verb+dict+ constructor rather than the literal \verb+{...}+.  Then,  the formulated rule is satisfied automatically,  since the key strings actually \textit{are} used as variable names.

It is possible to nest such dictionaries as above, and add variables of many standard Python types or NumPy arrays. This allows to perform a unified saving in a comprehensive way. For example, the following dictionary can be saved using directly the \verb+save+ function:
\begin{minted}{python}
dict_to_save = dict(
    info="This computation is a demonstration of SOLAX",
    params=dict(N_bath=N_bath, U_impurity=U),
    basis_from_nn=basis_final,
    last_epochs=dict(
        epochs=np.array([14, 15, 16, 17, 18]),
        accuracies=np.array([9.596332e-01, 9.524729e-01, 
        	9.450301e-01, 9.506576e-01, 9.534478e-01])
    ),
    random_keys_after=rand_keys
)

sx.save(dict_to_save, "solax_big_save_")
\end{minted}
\noindent At the technical level, the \verb+save+ and \verb+load+ functions process SOLAX objects in the following way:
\begin{itemize}
\item the standard Python types are converted to the JSON format~\cite{json} which is saved and loaded using the standard Python \verb+json+ module; 
\item the underlying NumPy arrays are saved and loaded using the NumPy means (the \verb+RandomKeys+ class needs additionally conversion of JAX arrays to and from NumPy).
\end{itemize}
We deliberately refrained from using the well known \verb+pickle+ module from the Python standard library, since it has been shown to have safety flaws~\cite{pickle}.

\subsection{Saving/loading BasisClassifier objects}

The \verb+BasisClassifier+ class does not follow the standard saving/loading mechanism described above. Instead, it implements its own methods \verb+save_state+ and \verb+load_state+ based on the Orbax library~\cite{orbax}. Saving is performed in a straightforward way:
\begin{minted}{python}
classifier.save_state("solax_nn_")
\end{minted}
\paragraph{Important note!} As in the case of the global \verb+save+ function, the shown method creates the directory before the \verb+BasisClassifier+ object is saved there. If the directory already exists, it will be first erased together with its current content.

In order to reconstruct the saved \verb+BasisClassifier+ object, the user needs first to create and initialize a new \verb+BasisClassifier+ instance following the procedure described in Section~\ref{sec:basis_classifier}:
\begin{minted}{python}
loaded_nn = sx.BasisClassifier(nn_call_on_bits)

fake_key = sx.RandomKeys.fake_key()
loaded_nn.initialize(fake_key, basis_start, optimizer)
\end{minted}
\noindent Here we reused the \verb+nn_call_on_bits+ function defining the NN architecture, \verb+basis_start+ as a prototype \verb+Basis+ instance, and the \verb+optimizer+ object. Instead of using a properly constructed JAX random key, we obtained a ``fake'' key directly from the \verb+RandomKeys+ class (i.e. without creating an instance and making an iteration step). Such fake keys should not be applied in true randomization, but can be used e.g. to initialize a NN whose neuron weights and biases will be anyway overwritten by loading. The saved NN state can be now restored as
\begin{minted}{python}
loaded_nn.load_state("solax_nn_")
\end{minted}
\noindent The presented tools are complete to provide the user with the possibility to conveniently and efficiently save and load SOLAX computations.

\subsection{A note on randomization under GPU acceleration}

The JAX library and its derivatives like FLAX used here for the NN implementation, treat randomness in a deterministic way based on random keys. However, in GPU-accelerated computations some non-deterministic effects have been observed, see e.g. the discussion~\cite{jax_nondeterm_discussion}. In our computations, we observed deviations of the NN training accuracies when rerunning the script from scratch. This could be avoided by adding the following lines immediately after the imports:
\begin{minted}{python}
import os
os.environ['XLA_FLAGS']='--xla_gpu_deterministic_ops=true'
\end{minted}
\noindent In subsequent JAX versions a different approach might be needed or the problem might be completely resolved.

\section{Conclusions and Outlook\label{sec:conclusions}}

In this work, we have demonstrated the foundational capabilities of the SOLAX library for tackling complex fermionic  many-body quantum systems. The core components of SOLAX implemented as the \verb+Basis+, \verb+State+, \verb+OperatorTerm+, \verb+Operator+ and \verb+OperatorMatrix+ classes, provide a robust framework for encoding and manipulating bases of Slater determinants, constructing quantum states, and handling operators within the second quantization formalism. Through a detailed application to the Single Impurity Anderson Model (SIAM), we illustrated how the iterative extension of the basis set combined with diagonalization can be efficiently implemented using SOLAX.

When the basis set grows too large to handle using available computational resources, the SOLAX built-in neural network (NN) support offers a practical solution. This approach allows for an efficient approximation of quantum states by identifying and selecting the most significant Slater determinants, thereby reducing the computational burden. The presented SIAM results, together with the SOLAX-based computations for the paradigmatic N$_2$ molecule performed in Ref.~\cite{Schmerwitz2024_N2}, underscore the flexibility and power of SOLAX in dealing with large basis sets. The modular integration of state-of-the-art machine learning techniques into the SOLAX framework opens up new avenues for addressing larger and more complex quantum systems that were previously beyond reach with alternative methods.


Looking forward, the development of additional toolboxes within the SOLAX package will be closely aligned with advancing research projects in quantum many-body physics and quantum chemistry. This requires enabling seamless interfacing between SOLAX and existing computational codes for, e.g., density functional theory (DFT) and Hartree-Fock computations. For instance, in our N$_2$ study\cite{Schmerwitz2024_N2}, SOLAX was successfully interfaced with the Python-based GPAW\cite{GPAW2024} code, which demonstrates the potential for straightforward integration with other Python-compatible packages in future research.  

Furthermore, a planned enhancement to SOLAX is the inclusion of a comprehensive library of predefined operators to facilitate the construction of effective Hamiltonians. Many quantum many-body problems require models built from standard elements, such as hopping and approximative interaction operators. The former will be defined based on lattice geometry, hopping range, and the symmetry of the involved orbital degrees of freedom, while the (typically onsite) interaction operator will be parametrized using physical quantities like the Hubbard \(U\) and Hund’s coupling \(J\). By incorporating such fundamental building blocks into an operator library, SOLAX aims to simplify and standardize Hamiltonian construction for a wide range of applications in condensed matter, quantum chemistry, and even for pedagogical purposes in lectures on these advanced topics.  

A major feature which is currently under development is the computation of spectral functions. Spectral functions are not only essential for studying the excitation spectra of cluster models, but will also enable SOLAX to serve as an impurity solver within the framework of dynamical mean-field theory (DMFT)\cite{dmft1,dmft2,dmft3}. Conceptually closely related to DMFT applications, we plan to implement predefined functionalities tailored for so called \textit{embedding schemes}. These schemes aim to partition the full problem into a subset of orbitals which will be treated with full many-body rigor, while the rest is maintained at a static mean-field level. 

All of the proposed features will be implemented in a modular way within Python, ensuring flexibility and ease of use. Through these developments, we hope to create a vibrant user/developer community which brings together expertise from a wide range of fields and reflecting the wide spectrum of potential applications for SOLAX.

\section*{Acknowledgements}
PB thanks Fred Baptiste for his valuable Python lessons. We gratefully acknowledge the group of Hannes J\'onsson at the University of Iceland for their collaboration on the N$_2$ molecule~\cite{Schmerwitz2024_N2} which was the first application of SOLAX. We further thank Henri Menke, Paul Fadler and Max Kroesbergen for useful discussions. The authors gratefully acknowledge the scientific support and HPC resources provided by the Erlangen National High Performance Computing Center (NHR@FAU) of the Friedrich-Alexander-Universität Erlangen-Nürnberg (FAU). PB gratefully acknowledges the ARTEMIS funding via the QuantERA program of the European Union.

\begin{appendix}
\label{section:appendix}

\end{appendix}
\FloatBarrier

\bibliography{refs}

\end{document}